\PassOptionsToPackage{usenames,dvipsnames}{xcolor}
\documentclass[a4paper,11pt]{article}
\usepackage{jheppub} 

\usepackage{amsmath,amssymb}
\usepackage{graphicx,subcaption,multirow}
\usepackage{tabularx}
\usepackage{slashed}
\usepackage{units}
\usepackage[colorlinks=true,linkcolor=blue,citecolor=blue,urlcolor=blue]{hyperref}
\usepackage{xcolor}
\usepackage{float}
\usepackage[utf8]{inputenc}
\usepackage[normalem]{ulem}

\def \L {\mathcal{L}} 
\def \rbar {\overline{r}} 
\def \Rbar {\overline{R}} 

\newcommand{\beq}{\begin{eqnarray}} 
\newcommand{\eeq}{\end{eqnarray}}

\usepackage{marginnote}

\newcommand{\changed}[2]{%
{\color{red}\sout{#1}}%
{\color{blue}\uwave{#2}}%
}

\title{Understanding the Quantized Angular Momentum\\ of Rotating Q-balls}

\author[a]{Benjamin DeVries,}
\author[b]{Fabrizio Vassallo,}
\author[c]{and Christopher B. Verhaaren}

\affiliation[a]{Department of Mathematics, Brigham Young University, Provo, UT, 84602, USA}
\affiliation[b]{Department of Physics, Wisconsin IceCube Particle Astrophysics Center, University of Wisconsin, Madison, Wisconsin 53706, USA}
\affiliation[c]{Department of Physics and Astronomy, Brigham Young University, Provo, UT, 84602, USA}

\emailAdd{bgd1729@student.byu.edu}
\emailAdd{fevassallo@wisc.edu}
\emailAdd{verhaaren@physics.byu.edu}

\abstract{Non-topological solitons, such as Q-balls, may contribute to the cosmological dark matter. The formation and evolution of Q-balls in the early universe requires an understanding of solitons with nonzero angular momentum. We derive (rather than assume) the schematic form of the scalar field configurations that produce rotating Q-balls, which produce their well known quantized angular momentum. This analysis leads to additional insight into the properties of these rotating solitons, including a method for computing their characteristic angular velocity. By considering rotating solitons in two spatial dimensions, we investigate these attributes concretely. We develop analytical approximations for the solitons and their defining quantities. We show that they agree with numerical results and exhibit the general properties of rotating solitons.}

\begin{document}
\maketitle
\flushbottom

\section{Introduction} \label{sec:introduction}
Since the pioneering efforts of~\cite{Friedberg:1976me,Friedberg:1977xf,Coleman:1985ki,Lee:1991ax}\changed{}{,} non-topological solitons have played a growing role in high-energy physics. Q-balls~\cite{Coleman:1985ki} are simple solitons, regularly used to explore a wide variety of phenomena. Applications include baryogenesis~\cite{Enqvist:1997si,Kasuya:1999wu,Multamaki:2002hv,Kasuya:2011ix,Tsumagari:2009na,vonHarling:2012yn}, inflation~\cite{Matsuda:2003gt,Matsuda:2004qi,Lloyd-Stubbs:2021xlk}, cosmological phase transitions~\cite{Kusenko:1997hj,Pearce:2012jp}, gravitational waves~\cite{Kusenko:2008zm,Croon:2019rqu,Wang:2021rfk,White:2021hwi,Hong:2024uxl,Bai:2024pki}, and models of alpha-clustered nuclei~\cite{Misicu:2018scg,Satarov:2021mli}.

Perhaps the most popular application is to suggest Q-balls contribute to the cosmological dark matter. Early explorations in supersymmetric theories~\cite{Kusenko:1997zq,Kusenko:1997si,Kusenko:1997vp,Dvali:1997qv,Kusenko:2004yw,Kusenko:2005du} and simple extensions of the Standard Model~\cite{Demir:1998zi} led to a variety of dark matter possibilities. These include self-interacting dark matter~\cite{Kusenko:2001vu,Enqvist:2001jd}, decaying dark matter~\cite{Kasuya:2024ldq}, macroscopic dark matter~\cite{Ponton:2019hux,Bai:2021mzu,Ansari:2023cay,Kamada:2025mji}, models of dark matter halos~\cite{Mielke:2002bp,Pombo:2025xsv}, and larger astrophysical objects~\cite{Troitsky:2015mda,Bai:2023mfi,DelGrosso:2024wmy,Kim:2025gck,SinghBhandari:2025ssx} including primordial black holes~\cite{Cotner:2016cvr,Cotner:2017tir,Cotner:2018vug,Hasegawa:2018yuy,Cotner:2019ykd,Flores:2021tmc,Flores:2021jas,Kasuya:2025puk}.

There are many ways to produce non-topological solitons in the early universe~\cite{Griest:1989bq,Postma:2001ea,Pearce:2022ovj,Bai:2022kxq,Jiang:2024zrb,Libanov:2024qfr,Azatov:2024npx}. Some assume miniclusters of scalar field with nonzero angular momentum can form Q-balls. Therefore, an understanding of rotating solitons is essential to comprehending the wealth of phenomena associated with cosmological solitons.

The analysis of rotating Q-balls began with their gravitational generalizations: boson stars. In 1994 an axisymmetric perturbative analysis~\cite{Kobayashi:1994qi} indicated that boson stars could not rotate slowly. This result is often assumed to also apply to the zero gravitation limit of Q-balls. As emphasized in~\cite{Almumin:2023wwi}, however, the axisymmetric scalar field used in the boson star analysis cannot carry angular momentum, so it is not clear that the boson star results can be applied to Q-balls.

The numerical simulation of boson stars~\cite{Schunck1996,Schunck:1996he} that followed the perturbative analysis assumed a scalar field of the form
\beq
\phi(t,r,\theta,\varphi)=f(t,r,\theta)e^{iN\varphi}~,\label{eq:BSansatz}
\eeq
for some integer $N$. The authors showed that any field configuration of this form leads to a relationship between the conserved particle number $Q$ and the angular momentum $J$,
\beq
J=NQ~,\label{eq:QuantJ}
\eeq
which they confirmed numerically. (While this result applies specifically to boson stars, other solitons can exhibit a similar relation between angular momentum and charge~\cite{Radu:2008pp, Radu:2008ta}.) With few exceptions (for instance~\cite{Kling:2020xjj}), subsequent analyses of rotating boson stars took~\eqref{eq:BSansatz} as their starting point~\cite{Silveira:1995dh,Kleihaus:2005me,Kleihaus:2007vk,Hartmann:2010pm,Kleihaus:2011sx,Liebling:2012fv,Davidson:2016uok,Herdeiro:2019mbz,Collodel:2019ohy,Delgado:2020udb,Dmitriev:2021utv,Gervalle:2022fze,Siemonsen:2023hko}. That their results agree with~\eqref{eq:QuantJ} numerically shows consistency with their starting assumptions, but does not demonstrate that quantized angular momentum is required. 

In agreement with the work on boson stars, most analyses of rotating solitons~\cite{Volkov:2002aj,Brihaye:2007tn,Brihaye:2008cg,Brihaye:2009yr,Campanelli:2009su,Brihaye:2009dx,Shnir:2011gr,Brihaye:2012uw,Radu:2012yx,Khaidukov:2013uia,Brihaye:2013ita,Kleihaus:2013tba,Nugaev:2014iva,Herdeiro:2014pka,Loiko:2018mhb,Loiko:2019gwk,Loginov:2020lwg,Loiko:2020htk,Blazquez-Salcedo:2022kaw,Saffin:2022tub,Zhang:2024ufh,Zhou:2024mea,Pombo:2025xsv,Brumelot2026}, including those considering 2+1 dimensions~\cite{Volkov:2002aj,Arodz:2009ye,Galushkina:2025yce,Galushkina:2025hur,Ivashkin:2025qdu}, also assumed a scalar field with the form given in~\eqref{eq:BSansatz}. A recent work~\cite{Almumin:2023wwi} examined this assumption and considered the possibility of Q-balls with small angular momentum. While they showed that such solitons could persist for long times (which may be sufficient for cosmological production), they did not construct an exact, stable soliton solution with small angular momentum.

In this work we elucidate the origin and implications of the field parameterization in~\eqref{eq:BSansatz}. Considering Q-balls in 2+1 and 3+1 dimensions, we derive~\eqref{eq:BSansatz} from specific assumptions about the soliton. This makes explicit which assumptions must be relaxed in order to obtain solitons that do not satisfy the relationship between angular momentum and charge in~\eqref{eq:QuantJ}.

For clarity, we refer to the 3+1 solitons as Q-balls. Their rotating generalizations have hollow centers, stabilized by centrifugal effects. We refer to them as Q-shells, similar to the solitons of gauged U(1) theories~\cite{Lee:1988ag, Gulamov:2015fya, Heeck:2021zvk} which are stabilized by repulsive gauge interactions~\cite{Arodz:2008nm,Ishihara:2021iag,Heeck:2021gam,Heeck:2021bce,Klimas:2022ghu,Satarov:2021mli}. We call the 2+1 dimensional solitons without angular momentum Q-disks, which have been explored in~\cite{Axenides:1999hs,Battye:2000qj}\textemdash and used to model certain condensed matter systems~\cite{Bunkov:2007fe}. In addition, we label their rotating generalizations Q-rings~\cite{Axenides:2001pi,Volkov:2002aj}, which are closely related to Q-tube solutions in 3+1 dimensions~\cite{Tamaki:2012yk,Nugaev:2014iva}.

Following the methods of~\cite{Heeck:2020bau}, we find approximate analytical descriptions of Q-disks and Q-rings as well as accurate numerical solutions. We focus on the lower-dimensional case because Q-rings are determined by an ordinary (rather than partial) differential equation. This makes Q-rings a simple, but nontrivial, system for exploring the details of rotating solitons. Our results confirm the general differential relation obtained in~\cite{Almumin:2023wwi},
\beq
dE=\omega dQ+\Omega dJ~.\label{eq:GenDiffRel}
\eeq
They also provide physical insight into the meaning of $\omega$ and $\Omega$, which as been overlooked in previous analyses. Our two-dimensional results also motivate further exploration of this differential relation in the more difficult, and especially interesting, three-dimensional case.

In the following section, Sec.~\ref{sec:non-rotating-Q-balls}, we review Q-balls and Q-disks. Much of our analysis is dimension-independent, so we highlight differences only when they are important. In either dimension, we find the standard ansatz follows from demanding a local minimization of an energy functional in which charge $Q$ is fixed. We generalize this procedure to rotating Q-shell and Q-rings in Sec.~\ref{sec:rotating-Q-balls}. This section connects the scalar field form in Eq.~\eqref{eq:BSansatz} to local minimization of the energy functional and assumptions about the character of the soliton. In Sec.~\ref{sec:Analytic} we develop analytic approximations for Q-disks and Q-rings. Comparing these approximate forms to numerical solutions in Sec.~\ref{sec:numerical-analysis}, we demonstrate their remarkably accuracy over much of the parameter space. This indicates that the analytic results can often be used instead of solving the systems numerically. We also outline a method for extracting the values of $\omega$ and $\Omega$ associated with a given soliton. After presenting our conclusions and directions for future work in Sec.~\ref{sec:conclusion}, we record a complete derivations of our analytical models for Q-disks and Q-rings in the Appendix~\ref{app:FullAnApprox}.

\section{Q-balls and Q-disks} \label{sec:non-rotating-Q-balls}
In this section we review the standard Q-ball construction. We do so, however, in a more systematic and general way than much of the literature. These methods produce the standard results, but also generalize in a useful way to the rotating solitons considered in the following section.

We begin with the Lagrangian density of a complex scalar field $\phi$ with a potential $U(|\phi|^2)$
\begin{align}
    \L = |\partial_\mu \phi|^2 - U(|\phi|^2)~, \label{eq:q-ball-lagrangian}
\end{align}
which produces the equations of motion
\beq
\ddot{\phi}-\nabla^2\phi+\phi\frac{dU}{d|\phi|^2}=0~,\label{eq:LagEoM}
\eeq
where $\partial_t\phi\equiv\dot{\phi}$.
As shown in~\cite{Coleman:1985ki}, this Lagrangian, which exhibits a global U(1) symmetry, leads to Q-ball solutions if the function $U(|\phi|^2)/|\phi|^2$ has a minimum at $|\phi| = \phi_0 / \sqrt{2}$  where $0 < \phi_0 < \infty$, and
\begin{align}
    0 \leq \sqrt{\frac{2}{\phi_0^2} U\left(\frac{\phi_0^2}{ 2}\right)} \equiv \omega_0 < m_\phi~, \label{eq:frequency-condition}
\end{align}
where the mass of the scalar field is defined to be
\beq
m_\phi^2\equiv\left.\frac{d^2U}{d\phi^\ast d\phi}\right|_{\phi=0}~.
\eeq

In this work, we consider solitons in both two and three spatial dimensions. So, following ~\cite{Tsumagari:2008bv}, we leave the number of dimensions arbitrary for the moment. For instance, the conserved charge associated with the continuous global symmetry is written as
\begin{align}
    Q = i \int d^Dx \left(\phi^*  \dot{\phi} - \phi  \dot{\phi}^*\right)~,\label{eq:Qdef}
\end{align}
where $D$ is the number of spatial dimensions.

We are interested in soliton solutions which minimize the energy for a fixed $Q$. The energy $E$ of any field configuration is given by 
\begin{align}
    E = \int d^Dx \left[|\dot{\phi}|^2 + |\nabla \phi|^2 + U(\phi^* \phi)\right].
\end{align}
To fix the charge, we introduce a Lagrange multiplier $\omega$ to enforce the above definition of $Q$
\begin{align}
    E(\omega) & = \int d^Dx \left[|\dot{\phi}|^2 + |\nabla \phi|^2 + U(\phi^* \phi)\right] + \omega \left[Q - i \int d^Dx \left(\phi^* \dot{\phi} - \phi \dot{\phi}^*\right)\right]. \label{eq:non-rot-energy-functional-phi}
\end{align}
The Lagrangian is related to this functional by
\begin{align}
    L =\omega Q-E(\omega)~.
\end{align}
It is then straightforward to show~\cite{Heeck:2020bau} the following relations:
\begin{align}
    \frac{dL}{d\omega} = Q~, \qquad & \frac{dE}{dQ} = \omega~.
\end{align}
The latter indicates that $\omega$, which was introduced as a Lagrange multiplier, can be interpreted physically as a chemical potential~\cite{Laine:1998rg,Nugaev:2019vru}. 

The variation of the energy functional in Eq.~\eqref{eq:non-rot-energy-functional-phi} with respect to $\phi$ and $\dot{\phi}$ leads to
\begin{align}
    \delta E(\omega) = & \int d^Dx\left[\left(-i\omega\dot{\phi} - \nabla^2\phi + \phi\frac{dU}{d|\phi|^2}\right)\delta\phi^\ast + \left(\dot{\phi} + i\omega\phi\right)\delta\dot{\phi}^\ast + \text{H.c.} \right]~.
\end{align}
Requiring the variation with respect to $\dot{\phi}$ to vanish leads to
\beq
\dot{\phi} + i \omega \phi = 0~. \label{eq:non-rotating-q-ball-constraint}
\eeq
This equation leads to what is often called the Q-ball ansatz. The variation with respect to $\phi$ produces
\begin{align}
    -i\omega\dot{\phi} - \nabla^2\phi + \phi\frac{dU}{d|\phi|^2} = 0 ~,\label{eq:NonRotEnVarPhi}
\end{align}
which is not quite the equation of motion from the Lagrangian given in~\eqref{eq:LagEoM}. If, however, we use the constraint in Eq.~\eqref{eq:non-rotating-q-ball-constraint} then both the relation in~\eqref{eq:NonRotEnVarPhi} and the Lagrangian equations produce
\beq
\nabla^2\phi =-\omega^2\phi+\phi\frac{dU}{d|\phi|^2}~.\label{eq:EoM}
\eeq
In other words, when subject to the constraint in Eq.~\eqref{eq:non-rotating-q-ball-constraint}, both the variation of the Lagrangian and the variation of the energy produce the same equation.

With this appreciation of the constraint in Eq.~\eqref{eq:non-rotating-q-ball-constraint}, we consider its solutions. These have the form
\begin{align}
    \phi = \frac{\phi_0}{\sqrt{2}} f(\vec{x}\,)e^{ig(\vec{x}\,) - i \omega t}~.\label{eq:NonRotfgParam}
\end{align}
Here $\vec{x}$ is the usual spatial position vector while $f(\vec{x})$ and $g(\vec{x})$ are functions from $\mathbb{R}^D$ into $\mathbb{R}$. The coefficient carries the dimensions of the scalar field in $\phi_0$ while the factor of $\sqrt{2}$ produces a convenient normalization. While this family of functions follows from the constraint in~\eqref{eq:non-rotating-q-ball-constraint}, we must also satisfy~\eqref{eq:EoM} to fully extremize the energy functional.

In particular, the equations of motion~\eqref{eq:EoM} induce two equations. The real part
\begin{align}
\nabla^2f=\frac{1}{\phi_0^2}\frac{dU}{df}+\left[\left(\nabla g \right)^2 -\omega^2\right]f~,\label{eq:fEqNoRot}
\end{align}
is the $f$ equation. The imaginary part
\begin{align}
    f\nabla^2g=-2\left( \nabla f\right)\cdot\left(\nabla g \right)~,\label{eq:gEqNoRot}
\end{align}
is the equation for $g$. In general, these equations are difficult to solve exactly. However, certain classes of solutions are physically motivated.

Inserting the form for the scalar field given in Eq.~\eqref{eq:NonRotfgParam} into the definition of $Q$~\eqref{eq:Qdef}, we find
\beq
Q=\omega\phi_0^2\int d^Dxf^2~.\label{eq:QdefF}
\eeq
Similar to the analysis in~\cite{Zhou:2024mea}, we find that the energy functional~\eqref{eq:non-rot-energy-functional-phi} takes the form
\begin{align}
    E(\omega) = \frac{1}{2} \omega Q + \int d^Dx \left[\frac{1}{2} \phi_0^2 (\nabla f)^2 +  \frac{1}{2} \phi_0^2 f^2 (\nabla g)^2 + U(f)\right]~.
\end{align}
Once again, we are interested in minimizing this quantity for a fixed value of $Q>0$. 

From the formula for $Q$~\eqref{eq:QdefF}, we see that $f(\vec{x})$ must be nontrivial for $Q$ to be nonzero. It must also be true that $f(\vec{x})$ goes to zero at spatial infinity so that $Q$ is finite. Therefore, the gradient of $f$ must be non-vanishing in at least some regions of space. The consequence of this is that, while both terms in $(\nabla f)^2 + f^2 (\nabla g)^2$ produce non-negative contributions to the energy density, only the former is required to be nonzero. 

This observation suggests we consider field configurations in which
\begin{equation}
\nabla g = 0~.
\end{equation}
The equations of motion for $f$~\eqref{eq:fEqNoRot} and $g$~\eqref{eq:gEqNoRot} indicate that this choice can produce nontrivial $f$ solutions. Furthermore, Eq.~\eqref{eq:QdefF} indicates that such nontrivial $f$'s can lead to nonzero, finite $Q$. Because any constant $g$ can simply be rephased away due to the U(1) symmetry of the theory, we can effectively choose $g=0$ when considering minimum energy soliton solutions.

What about the profile function $f$? The gradient of the $f$ term can be expressed as
\begin{align}
    (\nabla f)^2= &\left( \partial_r f\right)^2+\frac{1}{r^2}\left( \partial_\varphi f\right)^2,& &\text{2D}\\
   (\nabla f)^2=  &\left( \partial_r f\right)^2+\frac{1}{r^2}\left( \partial_\theta f\right)^2+\frac{1}{r^2\sin^2\theta}\left( \partial_\varphi f\right)^2.& &\text{3D}
\end{align}
Although we must have $\partial_r f \neq 0$ to keep $0 < Q < \infty$, $f$ need not depend on any angular coordinate, so choosing $f$ to be a function of only the radius further reduces the energy and is self-consistent with the $f$ equation~\eqref{eq:fEqNoRot} with $g=0$.

The components of the stress-energy tensor
\begin{align}
    T_{\mu\nu} = \partial_\mu \phi^* \partial_\nu \phi + \partial_\nu \phi^* \partial_\mu \phi - \eta_{\mu\nu} \L~, \label{eq:energy-momentum-tensor-definition}
\end{align}
provide another view of $f$ and $g$, as well as a check on our results. In our definition of $T_{\mu\nu}$, we have used the Minkowski metric $\eta_{\mu\nu}$ with negative spatial components. The $T^{0i}$ denote the momentum density along the $i$th spatial direction. Using the parameterization of the scalar field in~\eqref{eq:NonRotfgParam}, we find
\begin{align}
    T_{0r} &= -\phi_0^2 \omega f^2 \partial_r g&
    T_{0\varphi} &= -\phi_0^2 \omega f^2 \partial_\varphi g& & & &\text{2D}~, \\
    T_{0r} &= -\phi_0^2 \omega f^2 \partial_r g&
    T_{0\theta} &= -\phi_0^2 \omega f^2 \partial_\theta g& T_{0\varphi} &= -\phi_0^2 \omega f^2 \partial_\varphi g & &\text{3D}~.
\end{align}
By requiring $\partial_rg=0$ we select solitons with no radial momentum density. Similarly, choosing $g$ to be a constant ensures that they have no momentum density along the angular directions.

Similarly, we can relate the angular derivatives of $f$ to $T_{ij}$. These components of the stress-energy tensor correspond to the flux of $i$ momentum density along a surface of constant $j$. In other words, they represent shear stress density. Taking $g$ to be constant, we find
\begin{align}
    T_{r\varphi} &= \phi_0^2\partial_r f \partial_\varphi f& & & & & &\text{2D}~,\\
    T_{r\theta} &= \phi_0^2\partial_r f \partial_\theta f& T_{r\varphi} &= \phi_0^2\partial_r f \partial_\varphi f& T_{\theta\varphi} &= \phi_0^2\partial_\theta f \partial_\varphi f& &\text{3D}~.
\end{align}
Therefore, choosing $f$ to depend on $r$ only (for constant $g$) is equivalent to choosing a field with no shear stress densities.

Therefore, choosing $f(\vec{x}) = f(r)$ and $g(\vec{x}) = 0$ defines a minimum of the energy density with nontrivial $Q$ and $E$. We express this class of field configurations as
\begin{align}
    \phi = \frac{\phi_0}{\sqrt{2}} f(r) e^{-i \omega t}~,\label{eq:non-rotating-q-ball-solution}
\end{align}
which is the standard Q-ball ansatz. What we have shown above, however, is that the parameterization in Eq.~\eqref{eq:non-rotating-q-ball-solution} is more than a guess. It is the physically motivated field configuration that minimizes the energy of a nontrivial Q-ball system. 

Now that we know what the field parameterization in Eq.~\eqref{eq:non-rotating-q-ball-solution} means, we can use it to find specific soliton solutions. This form of the field leads directly to the Lagrangian
\beq
L = \int d^D x \left[\frac{1}{2} \phi_0^2 \omega^2 f^2 - \frac{1}{2} \phi_0^2 (\partial_r f)^2 - U(f)\right]~,
\eeq
and hence to an ordinary differential equation for the profile $f(r)$
\beq
\frac{d^2f}{dr^2}+\frac{D-1}{r}\frac{df}{df}+\omega^2f-\frac{1}{\phi_0^2}\frac{dU}{df}=0~.\label{eq:BallProfileEq}
\eeq
As is well known in three dimensions, a number of interesting properties of these solutions can be determined~\cite{Coleman:1985ki,Tsumagari:2008bv,Heeck:2020bau,Zhou:2024mea}, though we do not repeat them here. In Sec.~\ref{ssec:NonRotQDisks} we explore the  characteristics of related two-dimensional soliton solutions, Q-disks. In the following section, however, we extend the analysis of this section to solitons with nonzero angular momentum.

\section{Rotating Q-shells and Q-rings} \label{sec:rotating-Q-balls}
We use the same Lagrange multiplier method to require a field configuration has nonzero angular momentum and nonzero charge. Similar to above, one seeks the simplest non-trivial field configurations that minimize the energy $E$ subject to fixing nonzero values for both charge $Q$ and angular momentum $J$. We begin by recalling the definition of angular momentum in terms of the stress-energy tensor~\eqref{eq:energy-momentum-tensor-definition}
\begin{align}
    J^{ij} = \int d^D x \left(x^i T^{0j} - x^j T^{0i}\right)~.
\end{align}
For $D=2,3$, there is only one conserved angular momentum. In the $D=3$ case, we can always choose coordinates such that the angular momentum is along the $z$-axis. Therefore, in either case, we simply refer to a single angular momentum $J$ given by
\beq
J=-\int d^Dx~T_{0\varphi} = -\int d^Dx\left(\dot{\phi}^\ast\partial_\varphi\phi+ \dot{\phi}\partial_\varphi\phi^\ast\right)~,
\eeq
where $\varphi$ is taken to be the azimuthal angle of the polar coordinates of two dimensions and in the spherical polar coordinates of three dimensions.

Similar to the previous section, we define an energy functional. In this case, we introduce two Lagrange multipliers $\omega$ and $\Omega$, which relate to the charge $Q$ and angular momentum $J$, respectively. This functional has the form
\begin{align}
    E(\omega,\Omega) = & \int d^Dx \left[|\dot{\phi}|^2 + |\nabla \phi|^2 + U(\phi^* \phi)\right] + \omega \left[Q - i \int d^Dx \left(\phi^* \dot{\phi} - \phi \dot{\phi}^*\right)\right] \nonumber \\ 
    & + \Omega \left[J + \int d^D x \left(\dot{\phi}^\ast\partial_\varphi\phi+ \dot{\phi}\partial_\varphi\phi^\ast\right)\right]~. \label{eq:rotating-Q-ring-energyFuncPhi}
\end{align}
As shown in~\cite{Almumin:2023wwi}, we can write the Lagrangian in terms of this functional
\beq
L=\omega Q+\Omega J-E(\omega,\Omega)~.
\eeq
One then finds 
\begin{align}
\frac{dL}{d\omega}=Q~, \qquad \frac{dL}{d\Omega}=J~,
\end{align}
as well as the thermodynamic-like relation
\beq
dE=\omega dQ+\Omega dJ~.\label{eq:GenDifRel}
\eeq
As in the non-rotating case, $\omega$ is interpreted as a chemical potential. We also see that $\Omega$ behaves like a conjugate quantity to the angular momentum. In the case of a rigid body (or a black hole) we might call it the angular velocity, and it does indeed appear in Euler-like equations for the field's angular momentum~\cite{Almumin:2023wwi}. In this work, we simply refer to it as the characteristic angular velocity of the field.

When we vary the energy functional~\eqref{eq:rotating-Q-ring-energyFuncPhi} with respect to $\phi$ and $\phi^*$, we find 
\begin{align}
    \delta E(\omega, \Omega) =  \int d^D x \bigg[&\left(-i \omega \dot{\phi} - \nabla^2 \phi + \phi \frac{dU}{d|\phi|^2} -  \Omega  \partial_\varphi\dot{\phi}\right) \delta \phi^\ast  \nonumber \\
    &+ \left(\dot{\phi} + i \omega \phi + \Omega \partial_\varphi \phi\right) \delta\dot{\phi}^\ast + \text{H.c.} \bigg]~.
\end{align}
As in Sec.~\ref{sec:non-rotating-Q-balls}, requiring that the variation vanishes leads to two equations. The equation from the variation of $\dot{\phi}$ is
\beq
\dot{\phi}+\Omega\partial_\varphi\phi=-i\omega\phi~.\label{eq:RotConst}
\eeq
This relation was suggested and explored in~\cite{Almumin:2023wwi}, but here we see its true origin. The variation with respect to $\phi$ produces
\begin{align}
    i \omega \dot{\phi} + \nabla^2 \phi = \phi \frac{dU}{d|\phi|^2} - \Omega\partial_\varphi \dot{\phi}~.
\end{align}
Like in the nonrotating case, this equation agrees with the Lagrangian equations of motion~\eqref{eq:LagEoM} when the constraint equation~\eqref{eq:RotConst} is applied to both. In either case, one finds
\beq
\nabla^2\phi-2i\omega\Omega\partial_\varphi\phi-\Omega^2\partial_\varphi^2\phi=-\omega^2\phi+\phi\frac{dU}{d|\phi|^2}~,\label{eq:RotEoM}
\eeq
which leads to the $f$ and $g$ equations
\begin{align}
    \nabla^2f-\Omega^2\partial_\varphi^2f=&\frac{1}{\phi_0^2}\frac{dU}{df}+\left[\left(\nabla g\right)^2-\omega^2-2\omega\Omega\partial_\varphi g \right]f\label{eq:fEqRot}\\
    f\left[\nabla^2g-\Omega^2\partial_\varphi^2g\right]=&-2\left( \nabla f\right)\cdot\left( \nabla g\right)+2\Omega^2\left( \partial_\varphi f\right)\left( \partial_\varphi g\right)-2\omega\Omega\partial_\varphi f~. \label{eq:gEqRot}
\end{align}

Turning to the constraint equation~\eqref{eq:RotConst}, we note that the left-hand side looks like a derivative with respect to a linear combination of $t$ and $\varphi$. This motivates the change of variables
\begin{align}
u=\frac12\left( t+\frac{1}{\Omega}\varphi \right)~, \qquad v=\frac12\left(t-\frac{1}{\Omega}\varphi \right)~.
\end{align}
In these variables, we find
\beq
\left(\partial_t+\Omega\partial_\varphi \right)\phi=\partial_u\phi=-i\omega\phi~,
\eeq
which is solved by
\beq
\phi(r,\theta,u,v)=\frac{\phi_0}{\sqrt{2}}f(r,\theta,v)e^{i\widetilde{g}(r,\theta,v)-i\omega u}~.
\eeq
Here we have used three-dimensional coordinates. The two-dimensional case is obtained by simply omitting $\theta$. We expect to recover the non-rotating solitons discussed in the previous section in the limit of $\Omega\to0$. Thus, without loss of generality, we define the function $g(r,\theta,v)$ by
\beq
\widetilde{g}(r,\theta,v)=g(r,\theta,v)-\omega v~.
\eeq
Going back to $t$ and $\varphi$ coordinates, we have the following solution
\beq
\phi(t,r,\theta,\varphi)=\frac{\phi_0}{\sqrt{2}}f(r,\theta,\Omega t-\varphi)e^{ig(r,\theta,\Omega t-\varphi)-i\omega t}~.\label{eq:PhiGen_f_g}
\eeq
Similar to the analysis of Sec.~\ref{sec:non-rotating-Q-balls}, in order to minimize the energy functional the $f$ and $g$ functions must also satisfy the equations of motion \eqref{eq:fEqRot} and \eqref{eq:gEqRot}. We can, however, motivate certain functional forms by appealing to physical quantities.

Using this parameterization of $\phi$, we find the conserved quantities take the form
\begin{align}
    Q=&~\phi_0^2\int d^Dxf^2\left(\omega+\Omega\partial_\varphi g \right)~,\\
    J=&~\phi_0^2\Omega\int d^Dx\left[\left(\partial_\varphi f\right)^2+f^2\left(\partial_\varphi g\right)^2 \right]+\phi_0^2\omega\int d^Dxf^2\partial_\varphi g~,\\
    E=&~\int d^Dx\left\{\frac{\phi_0^2}{2}\left[\left(\partial_r f\right)^2+f^2\left(\partial_r g\right)^2 \right]+\frac{\phi_0^2}{2r^2}\left[\left(\partial_\theta f\right)^2+f^2\left(\partial_\theta g\right)^2 \right]\right.\nonumber\\
    &\qquad \qquad  +\frac{\phi_0^2}{2}\left(\frac{1}{r^2\sin^2\theta}+\Omega^2 \right)\left[\left(\partial_\varphi f\right)^2+f^2\left(\partial_\varphi g\right)^2 \right]+\omega\Omega\phi_0^2f^2\partial_\varphi g\nonumber\\
    &\qquad \qquad \left.+U(f)+\frac{\phi_0^2}{2}\omega^2f^2\right\}~.
\end{align}
Again, requiring that $0<Q<\infty$, we see that $f$ and $\partial_rf$ must be nonzero. We also see that the energy density is reduced by taking 
\beq
\partial_\theta g=\partial_r g=0~,\label{eq:ReduceEnCond}
\eeq
which is also consistent with the equations of motion~\eqref{eq:fEqRot}\textendash\eqref{eq:gEqRot}. In contrast to the non-rotating case, we cannot choose both $\partial_\varphi f=0$ and $\partial_\varphi g=0$ because we require a nonzero $J$. However, it is not yet clear which (or both) should be allowed to depend on $\varphi$.

We can better understand the way forward by considering the $g$ equation~\eqref{eq:gEqRot}. After enforcing the relations in~\eqref{eq:ReduceEnCond}, this becomes
\beq
\left(\frac{1}{r^2\sin^2\theta}-\Omega^2 \right)\left[2\left( \partial_\varphi f\right)\partial_\varphi g+f\partial_\varphi^2g \right]+2\omega \Omega\partial_\varphi f=0~.\label{eq:EoMImaginary}
\eeq
If we choose $\partial_\varphi g=0$, this equation is only satisfied for $\partial_\varphi f=0$, which would imply that $J$ is zero. We can, however, take
\beq
\partial_\varphi f=0~,\label{eq:RotFConst}
\eeq
to reduce the energy density while keeping $g$ nontrivial, so that $J\neq0$. When Eq.~\eqref{eq:RotFConst} is enforced the equation of motion given in Eq.~\eqref{eq:EoMImaginary} implies that
\beq
\partial_\varphi^2g=\partial_v^2g=0~.
\eeq
In other words, we find that
\beq
g(v)=g_0v=\frac{g_0}{2}\left(t-\frac{1}{\Omega}\varphi\right)~,
\eeq
where $g_0$ is some constant. 

The stress-energy tensor provides another view of this result. For general $f$ and $g$, the radial momentum density is
\beq
T_{0r}=-\phi_0^2\left\{ \omega\partial_rg+\Omega\left[ \left( \partial_rf\right)\partial_\varphi f+f^2\left( \partial_rg\right)\partial_\varphi g\right]\right\}~.
\eeq
Once we require $\partial_rg=0$, the only way to ensure that this momentum density vanishes is to have $\partial_\varphi f=0$. It is straightforward to show that this condition also ensures that the entire stress-energy tensor is independent of $\varphi$, because $\partial_\varphi^2g=0$. Though we cannot require $\partial_\varphi\phi=0$ and have nonzero angular momentum, this less restrictive quality of the stress-energy tensor reminds us of the axisymmetry we expect of systems rotating about a single axis.

What about $\partial_\theta f$? It is certainly true that if we choose $f$ to be independent of $\theta$ that we reduce the energy density. However, such a choice is inconsistent with the equations of motion for $f$, which (under the assumptions made above) now include the term
\beq
\frac{\left(\partial_\varphi g\right)^2}{r^2\sin^2\theta}f=\frac{g_0^2}{4\Omega^2r^2\sin^2\theta}f~.
\eeq
Because the equation for $f$ depends on $\theta$ explicitly, nontrivial $f$ solutions also depend on $\theta$.

In short, we have shown that the minimal energy field configurations with nonzero angular momentum take the form
\beq
\phi(t,r,\theta,\varphi)=\frac{\phi_0}{\sqrt{2}}f(r,\theta)e^{ig_0(t-\varphi/\Omega)/2-i\omega t}~.
\eeq
However, the field $\phi$ must also be single valued, so we require $\phi(\varphi+2\pi)=\phi(\varphi)$. Therefore, it must be that
\beq
-\frac{g_0}{2\Omega}=N~,
\eeq
for some integer $N$. This shows that the final form of the scalar field is
\beq
\phi(t,r,\theta,\varphi)=\frac{\phi_0}{\sqrt{2}}f(r,\theta)e^{-i\left(\omega+N\Omega \right)t+iN\varphi}~,\label{eq:RotFieldAnsatz}
\eeq
where the profile must satisfy
\beq
\frac{\partial^2f}{\partial r^2}+\frac{2}{r}\frac{\partial f}{\partial r}+\frac{1}{r^2}\frac{\partial^2f}{\partial \theta^2}+\frac{\cot\theta}{r^2}\frac{\partial f}{\partial\theta}+\frac{N^2}{r^2\sin^2\theta}f+\left(\omega+N\Omega\right)^2f-\frac{1}{\phi_0^2}\frac{dU}{df}=0~.\label{eq:ShellProfileEq}
\eeq
In two-dimensions there is no $\theta$ for the profile to depend on and the equation that determines $f$ is
\beq
\frac{\partial^2f}{\partial r^2}+\frac{1}{r}\frac{\partial f}{\partial r}+\frac{N^2}{r^2}f+\left(\omega+N\Omega\right)^2f-\frac{1}{\phi_0^2}\frac{dU}{df}=0~.\label{eq:RingProfileEq}
\eeq

Similar to the non-rotating solitons, the configuration in~\eqref{eq:RotFieldAnsatz} is not an ansatz. Rather, it results from seeking solutions that minimize the energy density for fixed, nonzero, charge $Q$ and angular momentum $J$. For this parameterization the charge is given by
\beq
Q=\phi_0^2\left(\omega+N\Omega \right)\int d^Dx\,f^2~,
\eeq
and the angular momentum produces the well-known, simple result of
\beq
J=NQ~.
\eeq
This is the famous quantization of angular momentum that arises in these classical soliton solutions. In contrast to previous work, this is an outcome of our analysis, not the result of an initial assumption.

Our derivation of this result can also provide a useful starting point for those interested in soliton solutions for which $J\neq NQ$. Any such solitons would follow from field configurations of the form~\eqref{eq:PhiGen_f_g}, with the functions $f$ and $g$ depending on more of the coordinates. It would be interesting to compare previous explorations~\cite{Kling:2017hjm,Almumin:2023wwi} of $J\neq NQ$ solitons to this form.

We also highlight that previous analyses of rotating Q-balls and boson stars typically express the combination $\omega+N\Omega$, which appears multiplying $t$ in~\eqref{eq:RotFieldAnsatz}, as simply $\omega$. This is reasonable because only this combination of parameters appears in the equation of motion. Our derivation makes clear, however,  that this combination is actually composed of two distinct, physical parameters: the chemical potential $\omega$ and the characteristic angular velocity $\Omega$. As we show in Sec.~\ref{sec:numerical-analysis}, the value of these two Lagrange multipliers can be determined for each soliton solution by leveraging the differential relation in~\eqref{eq:GenDifRel}.
 
When we insert $J=NQ$ into~\eqref{eq:GenDifRel}, we find
 \beq
 dE=\omega dQ+\Omega d(NQ)=\left(\omega+N\Omega \right)dQ+\Omega QdN~. \label{eq:general-differential-equation-with-dQ-and-dN}
 \eeq
For a family of solutions with fixed $N$, the last term vanishes and $\omega+N\Omega$ acts as the chemical potential\textemdash this parameter's previous interpretation. In contrast, by considering solutions at fixed $Q$ for various $N$, one can extract (at least formally) the characteristic angular velocity
 \beq
 \Omega = \frac{1}{Q}\left.\frac{dE}{dN}\right|_{Q}~.
 \eeq
 Similarly, writing $Q=J/N$, we find the relation
 \beq
 dE=\frac{1}{N}\left(\omega+N\Omega\right)dJ-\frac{\omega J}{N^2}dN~,
 \eeq
which demonstrates, at fixed $J$, that the true chemical potential is
 \beq
 \omega = -\frac{N^2}{J}\left.\frac{dE}{dN}\right|_{J}~.
 \eeq
In Sec.~\ref{sec:numerical-analysis}, we show that these relations are verified for rotating Q-rings in two dimensions. 
This indicates that $\omega$ and $\Omega$ describe physical quantities for each soliton, even though the equations that determine them depend only on $\omega+N\Omega$.

\section{Analytical Analysis\label{sec:Analytic}}
As indicated in the introduction, our analytical and numerical analyses focus on Q-disks and Q-rings. The two-dimensional Q-rings are determined by ordinary differential equations, making them a simple, yet nontrivial, system for exploring rotating solitons. We expect many of the qualitative results also arise in the very interesting three-dimensional generalization. Considering both the non-rotating and rotating solitons, we derive approximate analytical solutions for the profile, the charge, and the energy similar to~\cite{Ioannidou:2003ev,Ioannidou:2004vr, Heeck:2020bau}. Many of the same principles and qualities generalize to three or higher spatial dimensions. The 2D solutions are also similar to Q-tubes~\cite{Volkov:2002aj,Tamaki:2012yk,Nugaev:2014iva} or Q-strings~\cite{Chen:2024axd}, which are infinitely extended three-dimensional field configurations. In particular, the Q-ring equations coincide with the transverse cross-section equations of Q-tubes with nonzero winding number. We expect our methods generalize simply to describe these objects.

We derive novel analytical models of these objects using techniques similar to those employed in~\cite{Heeck:2020bau}. This section focuses on the most important and useful conclusions, with more detail reserved for the Appendix~\ref{app:FullAnApprox}.  In Sec.~\ref{sec:numerical-analysis} we compare these analytical models to numerical results.

\subsection{Q-disks\label{ssec:NonRotQDisks}}
We first consider Q-disks without angular momentum. 
Substitution of the non-rotating Q-ball solution \eqref{eq:non-rotating-q-ball-solution} into the equations of motion \eqref{eq:BallProfileEq} yields the following ordinary differential equation for the profile
\begin{equation}
    f''+\frac{1}{r}f'+\omega^2f-\frac{1}{\phi_0^2}\frac{dU}{df} = 0~.
\end{equation}

In this work we choose the potential $U(f)$ to be the sextic potential
\beq
U(\phi^\ast\phi)=m_\phi^2|\phi|^2-m_\phi\beta|\phi|^4+\xi|\phi|^6~.
\eeq
While this potential does not span the full possibilities of Q-balls, we expect qualitative results to hold generally~\cite{Heeck:2022iky,Almumin:2021gax}, although alternative calculational methods~\cite{Espinosa:2023osv} may be required.  Analytical solutions have also been obtained for other scenarios, such as the Q-tubes~\cite{Nugaev:2014iva} and Q-balls~\cite{Gulamov:2013ema} that arise from piecewise parabolic potentials.

Using the definitions of Sec.~\ref{sec:non-rotating-Q-balls}, we write $U(f)$ as
\begin{align}
    \frac{1}{\phi_0^2} U(f) = \frac{1}{2}(m_\phi^2 - \omega_0^2)(1-f^2)^2f^2 + \frac{1}{2} \omega_0^2 f^2~,
\end{align}
where
\beq
\phi_0=\sqrt{m_\phi\frac{\beta}{\xi}}~, \ \ \ \ \omega_0=m_\phi\sqrt{1-\frac{\beta^2}{4\xi^2}}~.
\eeq
This form motivates our using the dimensionless potential
\begin{align}
    V(f) \equiv \frac{1}{m_\phi^2 - \omega_0^2} \left(\frac{1}{2} \omega^2 f^2 - \frac{1}{\phi_0^2} U(f)\right) = \frac{1}{2} f^2 \left[\kappa^2 - (1-f^2)^2\right]~, \label{eq:dimensionless-potential-V}
\end{align}
where $\kappa$ is a dimensionless quantity defined by
\begin{align}
    \kappa^2 \equiv \frac{\omega^2 - \omega_0^2}{m_\phi^2 - \omega_0^2}~.
\end{align}
We also define the dimensionless radial coordinate
\begin{align}
    \rbar = r \sqrt{m_\phi^2 - \omega_0^2}~.
\end{align}
Throughout this paper we use a bar to indicate the dimensionless version of a quantity. For example, the dimensionless radius of a Q-disk is
\beq
\Rbar = R \sqrt{m_\phi^2 - \omega_0^2}~.
\eeq
We also find
\begin{align}
    Q & = \frac{2 \pi \omega \phi_0^2}{m_\phi^2 - \omega_0^2} \int d\rbar \, \rbar f^2~, \label{eq:non-rotating-q-disk-charge} \\
    E & = \omega Q + \pi \phi_0^2 \int d\rbar \, \rbar f^{\prime\,2}~. \label{eq:non-rotating-q-disk-energy}
\end{align}
In the energy relation we have used the virial theorem result
\beq
\omega Q=\frac{4\pi}{m_\phi^2 - \omega_0^2}\int d\overline{r}\,\overline{r}\,U(f)~.\label{eq:Virial}
\eeq

\begin{figure}[H]
    \centering    \includegraphics[width=0.505\textwidth]{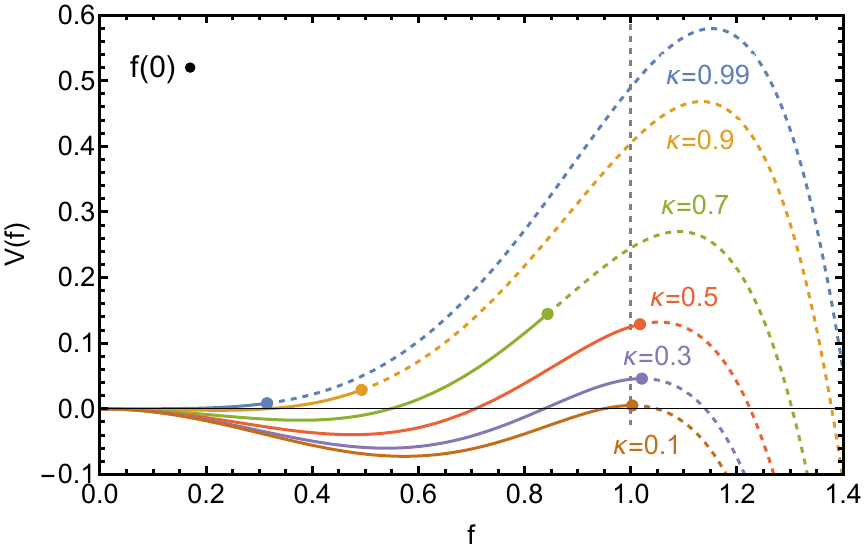}\hspace{0.1cm}\includegraphics[width=0.475\textwidth]{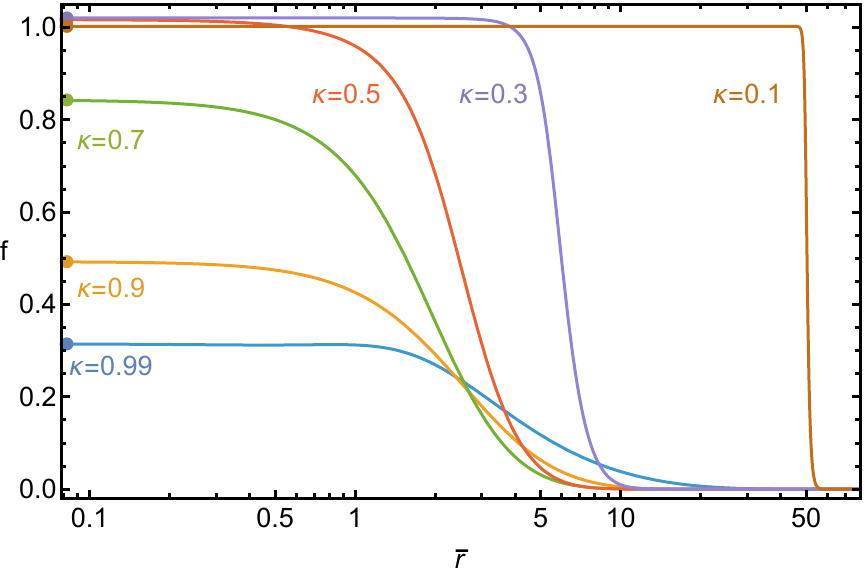}
    \caption{{\bf Left:} Plot of the effective potential $V(f)$ for several values of $\kappa$ and the particle trajectory related to the soliton solution. {\bf Right:} Q-disk profiles for the corresponding $\kappa$ values.}
    \label{fig:potential}
\end{figure}

Using $V(f)$ and $\rbar$ in the equations of motion, we arrive at
\begin{align}
    f'' + \frac{1}{\rbar}f' + \frac{dV}{df} = 0~. \label{eq:non-rotating-q-disk-equation-of-motion}
\end{align}
As with Q-balls, this has the form (by relating $f$ to the position of the particle and $\rbar$ to time) of an equation describing a particle in a potential with time-dependent friction. We use this analogy with single particle dynamics to guide our intuition about Q-disk profiles. To begin, we examine the effective potential $V(f)$ (see Fig.~\ref{fig:potential} for various examples). The relevant extrema of $V(f)$ are $f = 0$ and
\begin{align}
    f_\pm^2 = \frac{1}{3} \left(2 \pm \sqrt{1 + 3 \kappa^2}\right)~. \label{eq:non-rotating-potential-extrema}
\end{align}

To keep Eq.~\eqref{eq:non-rotating-q-disk-equation-of-motion}  finite at $\overline{r}=0$ we require $f'(\,\rbar \to0) = 0$. We also require $f(\rbar \rightarrow \infty) = 0$ to obtain a localized solution with a finite charge $Q$ and energy $E$. Within the single particle analogy, the second term in Eq.~\eqref{eq:non-rotating-q-disk-equation-of-motion} is a source of friction. Multiplying the equation by $f'$, integrating over position, and using the boundary conditions, produces 
\begin{equation}
    V(f(0)) = \int_0^\infty d\rbar\,\frac{f'\,^2}{\rbar}~,\label{eq:nonRotFriction}
\end{equation}
which we interpret (within the analogy) as the energy lost by the particle due to friction.

Friction plays an essential role in the trajectories of interest. It dissipates the initial potential energy so the particle can come to rest at $f = 0$ as $\rbar \rightarrow \infty$. As shown in the left panel of Fig.~\ref{fig:potential}, the potential maximum $V_\text{max} \gtrsim 0$ is small for small values of $\kappa$. Consequently, the particle must begin rolling from $f \approx f_+$ to ensure significant motion does not occur until large $\rbar$, when the frictional term is suppressed. For large values of $\kappa$, friction can only dissipate the initial potential energy if the particle starts well below $f_+$ and the transition happens earlier, meaning at smaller $\rbar$. These particle trajectories correspond to Q-disk profiles, which are shown in the right panel of Fig.~\ref{fig:potential}.

By adapting the methods used to determine Q-ball solutions\cite{Heeck:2020bau}, we obtain Q-disks. We find a \emph{transition function} that joins the interior and exterior solutions. These transition functions capture much of the essential  physics of global Q-balls, including radial excitations~\cite{Almumin:2021gax} and Q-balls in anti-de Sitter space~\cite{Rajaraman:2023ygy}. They can also elucidate gauged~\cite{Heeck:2021zvk,Heeck:2021gam} and Proca~\cite{Heeck:2021bce} Q-balls. We show below that transition functions characterize both Q-disks and Q-rings. We carefully outline how to derive these functions in the following subsection. While many Q-disk profiles are surprisingly well-approximated by transition functions, we also include a full profile approximation in Appendix \ref{appendix:non-rotating-analytical-model}.

\subsubsection{Transition Region} \label{sec:non-rotating-transition-function}
To derive the transition function we focus on the form of the function near the radius $\rbar \approx \Rbar$. Under the change of coordinates $x = \rbar - \Rbar$, the equation of motion becomes
\begin{gather}
	\frac{d^2 f}{dx^2} + \frac{1}{x + \Rbar} \frac{df}{dx} + \frac{dV}{df} = 0~.
\end{gather}
For $\overline{R}\gg1$ (often referred to as the thin-wall limit), the friction term is subdominant. Dropping this term, we find
\begin{align}
	\frac{d^2 f_t}{dx^2} = f_t\left[  \left(1 - f_t^2\right) \left(1 - 3f_t^2\right)-\kappa^2\right]~,\label{eq:TransEqWithKap}
\end{align}
where $f_t$ is the transition function. This equation can be rewritten as
\begin{align}
    \frac{d}{dx} \left[\frac{1}{2} \left(\frac{df_t}{dx}\right)^2 - \frac{1}{2} f_t^2 \left(1 - f_t^2\right)^2+\kappa^2\frac{f_t^2}{2}\right] = 0 ~,
\end{align}
which implies that
\beq
\left(\frac{df_t}{dx}\right)^2=f_t^2 \left(1 - f_t^2\right)^2-\kappa^2f_t^2+c_0~,
\eeq
where $c_0$ is some constant. Inserting this into the relation for the energy loss due to friction~\eqref{eq:nonRotFriction}, we find
\beq
V(f(0))=\int_{-\overline{R}}^\infty\frac{dx}{x+\overline{R}}f_t^2\left[ \left(1-f_t^2\right)^2-\kappa^2\right]+c_0\int_{-\overline{R}}^\infty\frac{dx}{x+\overline{R}}~.
\eeq
The last integral is infinite, which implies that $c_0=0$ and so that
\beq
\left(\frac{1}{f_t}\frac{df_t}{dx}\right)^2=\left(1 - f_t^2\right)^2-\kappa^2~.
\eeq

This equation is difficult to solve for general $\kappa$. However, we argued above that large $\overline{R}$ corresponds to small $\kappa$. This leads us to consider the $\kappa=0$ case, for which
\beq
 \frac{d f_t}{dx} = \pm f_t(1 - f_t^2)~.
\eeq
We integrate this to obtain the function
\begin{align}
    f_t(\rbar) = \left[1 + c_t e^{\mp2(\rbar - \Rbar)}\right]^{-1/2}~.\label{eq:UpDownTransition}
\end{align}

To ensure a finite energy solution, which starts at a positive value at $\overline{r}<\overline{R}$ and then goes to zero for $\overline{r}>\overline{R}$, we choose the positive sign in the exponent. The coefficient $c_t$ is determined by defining the radius $\Rbar$ by $f''(\Rbar) = 0$. This results in the transition function
\begin{align}
    f_t (\rbar) = \left[1 + 2 e^{2(\rbar - \Rbar)}\right]^{-1/2}~.
\end{align}

Putting this into the relation for the energy lost to friction we find
\beq
V(f(0))=4\int_{-\overline{R}}^\infty\frac{dx}{x+\overline{R}}\frac{e^{4x}}{\left(1+2e^{2x}\right)^3}~.
\eeq
This integrand is sharply peaked at $x=0$, so it is well approximated by
\begin{align}
V(f(0))&\approx\frac{4}{~\overline{R}~}\int_{-\overline{R}}^\infty\frac{e^{4x}}{\left(1+2e^{2x}\right)^3}+\mathcal{O}\left(\overline{R}^{-2}\right)\nonumber\\
&=\frac{4}{~\overline{R}~}\left(\frac{1}{16}-\frac{e^{-4\overline{R}}}{4\left(1+2e^{-2\overline{R}} \right)^2} \right)+\mathcal{O}\left(\overline{R}^{-2}\right)\nonumber\\
&\approx\frac{1}{4\overline{R}}~.
\end{align}
In the $\kappa=0$ limit the maximum of the potential satisfies $V(1)=0$, so the particle trajectory must begin at the exact maximum. Consequently, in the infinitely-thin-wall limit the particle cannot lose any energy to friction as it rolls to the maximum at $V(0)=0$. In other words, for $\kappa=0$ it must be that $\overline{R}=\infty$.

A much more useful transition function is one that holds for a larger range of $\kappa$. We seek such a function by keeping the same functional form of $f_t$, but allow it to be rescaled by a $\kappa$-dependent coefficient. In short, we define a modified transition function
\beq
f_T=f_0(\kappa)f_t(\overline{r})~.
\eeq

By substituting this ansatz into the energy lost to friction relation we find
\begin{align}
    V(f(0))=\frac{f_0^2}{2}\left[\kappa^2-\left(1-f_0^2 \right)^2 \right]=4f_0^2\int_{-\overline{R}}^\infty\frac{dx}{x+\overline{R}}\frac{e^{4x}}{\left(1+2e^{2x}\right)^3}~.
\end{align}
This leads to 
\beq
\kappa^2\approx\frac{1}{2\overline{R}}+\left(1-f_0^2 \right)^2 ~.\label{eq:kapRf1}
\eeq
We can also insert this transition function into the equations of motion at $\overline{r}=\overline{R}$. Because we have defined the radius by $f''(\overline{R})=0$ we find
\beq
\kappa^2\approx\frac{3}{2\overline{R}}+\left(1-f_0^2 \right)\left(1-\frac13f_0^2 \right) ~,\label{eq:kapRf2}
\eeq
where we have used that $f_t(\overline{R})=1/\sqrt{3}$.

By using~\eqref{eq:kapRf1} and~\eqref{eq:kapRf2} we can eliminate $\overline{R}$ and find
\beq
\kappa^2=\left(1-f_0^2 \right)\left(1-3f_0^2 \right)~.
\eeq
This is exactly the equation satisfied by $f_{\pm}$, as defined in Eq.~\eqref{eq:non-rotating-potential-extrema}. We choose the $f_+$ solution as it has the correct limiting behavior as $\kappa\to0$. Substituting this result back into~\eqref{eq:kapRf1} we find
\begin{align}
    \Rbar = \frac{1}{4 f_+^2 (f_+^2 - 1)} = \frac{9}{4(3\kappa^2 + \sqrt{1 + 3\kappa^2} - 1)} \approx \frac{1}{2\kappa^2} + \frac{1}{8} - \frac{5 \kappa^2}{32} + \mathcal{O}(\kappa^4)~. \label{eq:R-bar-approximation-formula}
\end{align}
Note the approximate inverse (square) relationship between $\Rbar$ and $\kappa$. The analogous expression for Q-balls~\cite{Heeck:2020bau}, differs by a factor of 2, a result of the frictional terms in two dimensions versus three dimensions. So, while the functional form (in terms of $\overline{R}$) is the same, the dependence on $\kappa$ is different.

We emphasize that 
\beq
f_T(\overline{r})=\frac{f_+(\overline{R})}{\sqrt{1+2e^{2(\overline{r}-\overline{R})}}}~, \label{eq:Q-disk-transition-function}
\eeq
with
\beq
f_+^2=\frac12+\frac12\sqrt{1+\frac{1}{\,\overline{R}\,}}~,
\eeq
is not a solution of Eq.~\eqref{eq:TransEqWithKap}. However, it is a solution if we drop terms of order $\overline{R}^{-1}$. This is the approximation made of the full equations to obtain Eq.~\eqref{eq:TransEqWithKap}, so our transition function is a self-consistent approximation of the true solution in the large $\overline{R}$ limit.

We can approximate the charge and energy of a Q-disk by substituting the transition function into \eqref{eq:non-rotating-q-disk-charge} and \eqref{eq:non-rotating-q-disk-energy}. We find
\begin{align}
    Q&=\frac{\pi\omega\phi_0^2f_+^2}{m_\phi^2-\omega_0^2}\left[\left( \overline{R}-\ln\sqrt{2}\right)^2+\frac{\pi^2}{12}\right]+\mathcal{O}\left(e^{-2\overline{R}} \right)~, \label{eq:non-rotating-Q-approximation}\\
    E&=\omega Q+\frac{\pi\phi_0^2f_+^2}{4}\left( \overline{R}-\ln\sqrt{2}+\frac12\right)+\mathcal{O}\left(e^{-2\overline{R}} \right)~.\label{eq:non-rotating-E-approximation}
\end{align}
These results confirm that the charge and energy scale like the area of the disk, although an additional contribution to the energy scales like the circumference.

\begin{figure}[ht]
    \centering    
    \includegraphics[width=0.49\textwidth]{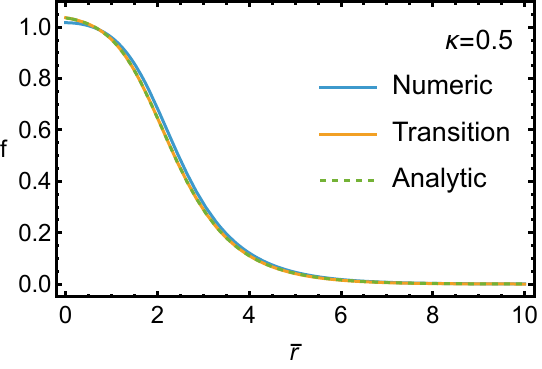}
\includegraphics[width=0.49\textwidth]{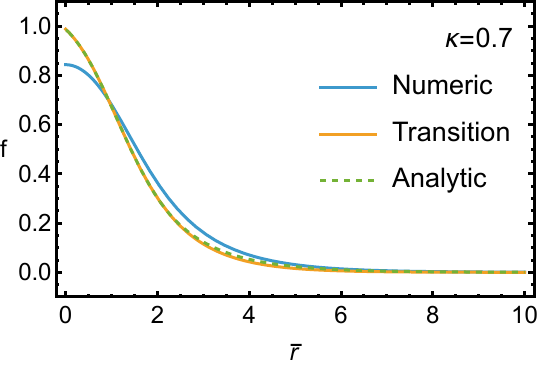}
    \caption{Comparison of the numerical solution (blue), transition function (orange), and full analytic approximation (dashed green) for $\kappa=0.5$ (left) and $\kappa=0.7$ (right).}
    \label{fig:num-full-trans-comparison-0.5}
\end{figure}

The transition function compares well with numerical solutions for the profile. In the left panel of Fig.~\ref{fig:num-full-trans-comparison-0.5} we see that the full analytical approximation (dashed green) approximates the numerical solution (blue) quite well for $\kappa=0.5$. We also see that the analytic approximation and the transition function lie nearly on top of each other. For larger $\kappa$ the agreement with the numerical solution is not as good but gives rough agreement, as shown in the left panel of the figure. The full analytic approximation is a moderate improvement of the transition function. Finding even rough agreement is remarkable away from the large radius limit for the transition function.

\begin{figure}[ht]
    \centering    
    \includegraphics[width=0.49\textwidth]{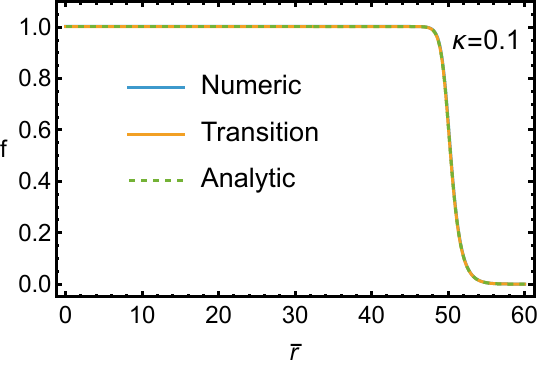}
\includegraphics[width=0.49\textwidth]{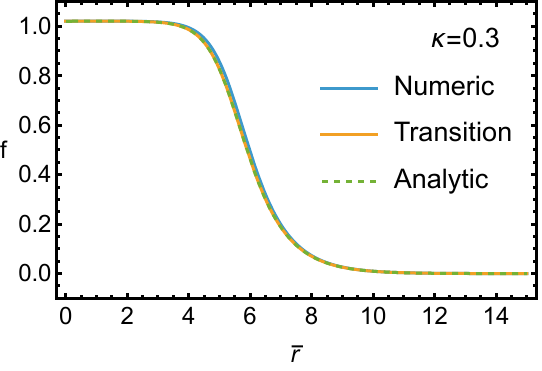}
    \caption{Comparison of the numerical solution (blue), transition function (orange), and full analytic approximation (dashed green) for $\kappa=0.1$ (left) and $\kappa=0.3$ (right).}
    \label{fig:num-trans-comparison-0.1}
\end{figure}

In Fig.~\ref{fig:num-trans-comparison-0.1} this comparison for $\kappa = 0.1$ (left) and $\kappa=0.3$ (right) demonstrates the accuracy of the transition profile in the large-$Q$, or thin-wall, limit. Equivalently, since the approximation is controlled by the inverse soliton size $\overline{R}^{-1}$, agreement is best when $\overline{R} \gg 1$. For $\kappa=0.3$ we see there is almost no difference between the numerical and analytical results.  Only zooming in on the transition reveals a modest discrepancy in the location of the radius. Shifting the radius used in the transition function slightly provides a very good fit to the numerical solution. The same hold true for smaller values, like $\kappa = 0.1$. For these larger-radius solitons the transition function (up to errors related to accurately estimating the radius) is an excellent approximation of the numerical results. Consequently, in the large-radius limit, the equations for the charge~\eqref{eq:non-rotating-Q-approximation} and energy~\eqref{eq:non-rotating-E-approximation} are close to their true values, as shown in Sec.~\ref{sec:numerical-analysis}.

\subsection{Q-rings\label{ssec:RotQRings}}
In this section we develop approximate analytical solutions for the profile, charge, angular momentum, and energy of rotating Q-disks. The methods largely parallel the non-rotating case, but include important differences. Extending these formulae to describe rotating Q-balls may provide novel insights into rotating solitons in 3 dimensions.

To begin, we write the differential equation~\eqref{eq:RingProfileEq}, which determines the profile $f$, in terms of the dimensionless coordinate $\rbar = r \sqrt{m^2 - \omega_0^2}$ and the dimensionless potential $V$ defined in \eqref{eq:dimensionless-potential-V}. In contrast to Q-disks, however, $\kappa$ now depends on $N$ and $\Omega$:
\begin{align}
    \kappa^2 \equiv \frac{\widetilde{\omega}^{\,2}-\omega_0^2}{m_\phi^2 - \omega_0^2}~, \qquad \widetilde{\omega} &\equiv \omega + N\Omega~. \label{eq:kappa-squared-and-omega-tilde-definitions}
\end{align}
Using this new definition of $\kappa$ profile is determined by
\begin{align}
    f'' + \frac{1}{\rbar} f' - \frac{N^2}{\rbar\,^2} f + \frac{dV}{df} = 0~, \label{eq:rotating-q-disk-equation-of-motion}
\end{align}
which differs from the non-rotating equation \eqref{eq:non-rotating-q-disk-equation-of-motion} by a single $N$-dependent term.

\begin{figure}[H]
    \centering    
    \includegraphics[width=0.505\textwidth]{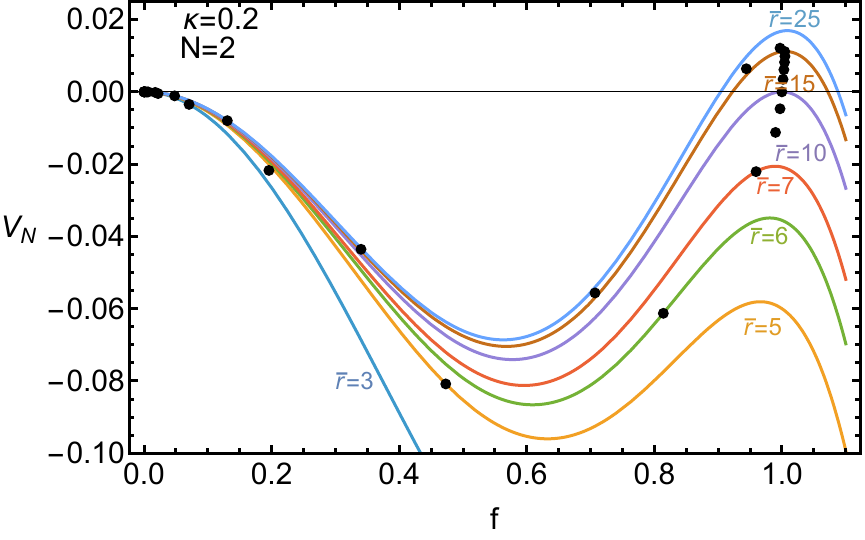}
    \includegraphics[width=0.475\textwidth]{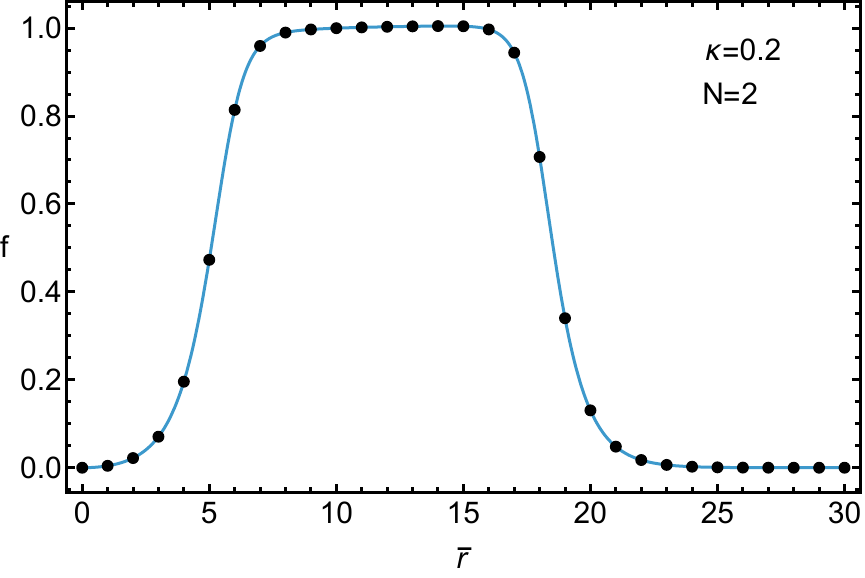}
    \caption{{\bf Left:} Plot of the potential $V_N(f,\rbar)$ at various values of $\rbar$ for $\kappa = 0.2$ and $N = 2$. {\bf Right:} Soliton profile for the same parameter values. The black dots denote the value of the soliton profile at steps of integer $\rbar$.}
    \label{fig:potential-with-n-over-time}
\end{figure}

Similar to the non-rotating analysis, \eqref{eq:rotating-q-disk-equation-of-motion} has the form of an equation describing the position of a particle in a potential with time-dependent friction. Unlike the non-rotating case, however, the total potential
\beq
V_N(f,\overline{r})=\frac{f^2}{2}\left[\kappa^2-\left(1-f^2 \right)^2-\frac{N^2}{\overline{r}^2} \right]~,
\eeq
depends on $\overline{r}$. Figure~\ref{fig:potential-with-n-over-time} illustrates how increasing $\rbar$ makes the potential approach the $N=0$ form.

This $\rbar$ dependence implies that in the particle trajectory analysis the potential is \emph{time-dependent}. While the (non-rotating) profile potential for gauged~\cite{Heeck:2021zvk} and Proca~\cite{Heeck:2021bce} Q-balls depends on $\overline{r}$ through the gauge field, this explicit dependence is qualitatively different. Moreover, while the corresponding potential for (non-rotating) Q-balls in anti-de Sitter space~\cite{Rajaraman:2023ygy} does exhibit this explicit dependence on the radial coordinate, the present dependence is quite different. As shown below, the Q-ring $\rbar$ dependence results in profiles that are more similar to Q-shells~\cite{Heeck:2021gam,Heeck:2021bce}.

As with Q-disks, the extrema of this potential (as a function of $f$) are particularly useful. These can be seen in the left panel of Fig~\ref{fig:potential-with-n-over-time}. We find a local maximum at $f=0$ along with a local minimum at $f_{-}(\kappa,\overline{r})$ and a maximum at $f_{+}(\kappa,\overline{r})$. The latter two are
\beq
f_{\pm}^2(\kappa,\overline{r})=\frac13\left(2\pm\sqrt{1+3\kappa^2-3\frac{N^2}{\overline{r}^2}} \right)~,
\eeq
which implies that these extrema only exist for
\beq
\rbar>N\sqrt{\frac{3}{1+3\kappa^2}}~.\label{eq:rConstraint}
\eeq
 In addition, for any $\kappa$, at the radius $\overline{r}=N/\kappa$ the profile and potential satisfy
 \beq
 f_+(\kappa,N/\kappa)=1~, \ \ \ \ V(f_+(\kappa,N/\kappa),N/\kappa)=0~.\label{eq:NoverKappa}
 \eeq
As discussed below, these results constrain possible particle trajectories that end at $f=0$.

The physical quantities that characterize Q-rings are defined in terms of the profile. It is straightforward to show that
\begin{align}
    Q & = \frac{2 \pi \widetilde{\omega} \phi_0^2}{m^2 - \omega_0^2} \int d\rbar \, \rbar f^2,\label{eq:rotating-Q-ring-charge}\\
    J &= NQ~, \\
    E &= \widetilde{\omega} Q + \pi \phi_0^2 \int d\rbar \, \rbar \left(f'\,^2 + \frac{N^2}{\rbar^2} f^2\right)~.\label{eq:rotating-Q-ring-energy}
\end{align}
In the energy relation we have used the virial relation, which matches Eq.~\eqref{eq:Virial} with $\omega\to\widetilde{\omega}$.

As above, we require the profile to vanish as $\overline{r}\to\infty$ so that the particle number $Q$ and energy $E$ are finite. We insert the power series expansion 
\begin{equation}
f(\rbar) = \sum_{n=0}^\infty a_n\rbar^n~,
\end{equation}
into \eqref{eq:rotating-q-disk-equation-of-motion} to find conditions on $f$ as $\overline{r}\to0$. The result 
\begin{align}
0=&-a_0\frac{N^2}{\rbar^2}+\frac{a_1}{\rbar}\left(1-N^2\right)+\sum_{n=0}^\infty r^n\left\{a_{n+2}\left[\left(n+2\right)^2-N^2 \right]+a_n(\kappa^2-1) \right\}\nonumber\\
&+4f^3-3f^5~,\label{eq:SmallRexpansion}
\end{align}
indicates that $a_0=0$ whenever $N>0$. This also ensures there are no singularities in the energy integrand~\eqref{eq:rotating-Q-ring-energy}. The second term in~\eqref{eq:SmallRexpansion} implies that $a_1=0$ for $N>1$. Consequently, if the first nonzero $a_n$ is denoted $a_i$, then for $N>1$ we find
\beq
a_{n+2}\left[ \left(n+2\right)^2-N^2 \right]=-a_n(\kappa^2-1) \ \ \ \ \text{for $n<3i$}~,
\eeq
because the $f^3$ and $f^5$ terms appear with higher powers of $\rbar$. This forces all $a_n$ with $n<N$ to be zero. In other words, the boundary condition as $\rbar\to0$ is
\beq
\lim_{\rbar\to0}f(\rbar)\propto \rbar^N~.
\eeq
In short, $f(0)=0$ and profiles for higher $N$ are even more suppressed near the origin. 

In general, Q-ring profiles increase from $f(0) = 0$ to some peak value then return to $f(\infty) = 0$. We note that $N=1$ Q-rings are somewhat distinct from higher $N$. While $f'(0)$ can be a nonzero constant for $N = 1$, the profiles at every subsequent value of $N$ satisfy $f'(0) = 0$. So, while the $N>1$ trajectories start from rest, the $N=1$ particle can have an initial velocity.

By multiplying the equation of motion~\eqref{eq:rotating-q-disk-equation-of-motion} by $f'$ and integrating over $\overline{r}$ we obtain the Q-ring energy loss formula. Because $f(0)=0$ and $V(0)=0$, this leads to
\beq
\frac12\left(f'(0)\right)^2=\int_0^\infty\frac{d\rbar}{\rbar}f'\left(f'-\frac{N^2}{\rbar}f \right)~.
\eeq
In the rolling particle analogy for Q-rings the trajectory begins and ends at $V(0)=0$. Therefore, the friction does not need to balance the initial potential energy. Instead, the potential's time dependence injects energy into the system that must be balanced by energy lost to friction. We also note the left-hand side of this equation vanishes, except for $N=1$ where the initial kinetic energy must also be lost to friction.

To summarize, the particle begins at the origin, with zero potential energy. Due to the initial shape of the potential (see the left panel of Fig.~\ref{fig:potential-with-n-over-time} for small $\rbar$ plots), the particle slides downhill to the right. If $N>1$ this transition is not immediate. As $\rbar$ increases so does the maximum at $f_+$. Eventually, the particle slows to a stop at some point with non-negative potential energy on the left side of $f_+$, then reverses direction. As the particle slides back toward the origin, its potential and kinetic energy are lost to friction, allowing it to comes to rest exactly $f=0$. The resulting profile is shown in the right panel of Fig.~\ref{fig:potential-with-n-over-time}.

As with Q-disks, the profile can be divided into interior, exterior, and transition regions. For Q-rings, however, the transition region is more complicated. Nevertheless, we can model this region effectively with the transition function derived in the non-rotating case.

\subsubsection{Transition Region} \label{subsec:rot-transition}
For small $\kappa$ the maximum possible value of the potential at $f_+$ is barely above zero. This means that the particle must roll back to the origin with very little friction. In other words, this transition must occur at large $\rbar$. In this large radius limit we found two possible profile solutions~\eqref{eq:UpDownTransition}. One transitions from zero to large values while the other transitions from large values to zero. This motivates us to approximate the profile as a product of two transition functions\textemdash one that starts at zero and transitions up followed by another that returns to zero:
\begin{align}
    f_t (\rbar) = \left[\left(1 + 2 e^{-2(\rbar - \Rbar_<)}\right)\left(1 + 2 e^{2(\rbar - \Rbar_>)}\right)\right]^{-1/2}~.
\end{align}

In this definition, the parameters $\Rbar_<$ and $\Rbar_>$ are positions (with $\Rbar_<<\Rbar_>$) such that $f_t^{\prime\prime}$ is close to zero. Specifically, we find
\beq
f''_t(\Rbar_{<,>})=0+\mathcal{O}\left(e^{-2(\Rbar_>-\Rbar_<)} \right)~,
\eeq
which indicates that when the two radii are sufficiently separated the two transitions are effectively independent. Profiles of this type are quite accurate (in three dimensions) for gauged~\cite{Heeck:2021gam} and Proca~\cite{Heeck:2021bce} Q-shells as well as for the shells around Q-balls that arise from radial excitations~\cite{Almumin:2021gax}.

As in the non-rotating case, we obtain a more accurate profile away from the $\kappa=0$ limit by simply multiplying by $f_+(\rbar,\kappa)$. That is, we use the transition function
\beq
f_T(\rbar)=\frac{f_+(\rbar)}{\sqrt{\left(1 + 2 e^{-2(\rbar - \Rbar_<)}\right)\left(1 + 2 e^{2(\rbar - \Rbar_>)}\right)}}~. \label{eq:Q-ring-transition-function}
\eeq
This immediately leads to
\beq
f_T(\Rbar_<)=\frac{f_+(\Rbar_<)}{\sqrt{3\left(1+2e^{-2\Delta\Rbar} \right)}}~, \ \ \ f_T(\Rbar_>)=\frac{f_+(\Rbar_>)}{\sqrt{3\left(1+2e^{-2\Delta\Rbar} \right)}}~,
\eeq
where we define the difference in the radii by
\beq
\Delta\Rbar\equiv\Rbar_>-\Rbar_<~.
\eeq
It is also useful to evaluate the first and second derivatives of these functions at each radius. The first derivatives of the transition function satisfy
\begin{align}
   \frac{ f'_T(\Rbar_<)}{f_T(\Rbar_<)}&=\frac{N^2}{2f_+(\Rbar_<)^2\Rbar_<^3\left(3f_+(\Rbar_<)^2-2 \right)}+3\frac{f_T(\Rbar_<)^2}{f_+(\Rbar_<)^2}-\frac13\\
    \frac{f'_T(\Rbar_>)}{f_T(\Rbar_>)}&=\frac{N^2}{2f_+(\Rbar_>)^2\Rbar_>^3\left(3f_+(\Rbar_>)^2-2 \right)}-3\frac{f_T(\Rbar_>)^2}{f_+(\Rbar_>)^2}+\frac13~.
\end{align}
The second derivatives are
\begin{align}
    \frac{f''_T(\Rbar_<)}{f_T(\Rbar_<)}=&\frac{N^2\left[3f_T(\Rbar_<)^2/f_+(\Rbar_<)^2-\frac13 \right]}{f_+(\Rbar_<)^2\Rbar_<^3\left( 3f_+(\Rbar_<)^2-2\right)} +\left(3\frac{f_T(\Rbar_<)^2}{f_+(\Rbar_<)^2}-1 \right)\left(9\frac{f_T(\Rbar_<)^2}{f_+(\Rbar_<)^2}+\frac13 \right) \nonumber \\
    &-\frac{N^4\left(9f_+(\Rbar_<)^2-2 \right)+6\Rbar_<^2N^2f_+(\Rbar_<)^2\left( 3f_+(\Rbar_<)^2-2\right)^2}{4f_+(\Rbar_<)^4\Rbar_<^6\left( 3f_+(\Rbar_<)^2-2\right)^3}~,\\
    \frac{f''_T(R_>)}{f_T(\Rbar_>)}=&-\frac{N^2\left[3f_T(\Rbar_>)^2/f_+(\Rbar_>)^2-\frac13 \right]}{f_+(\Rbar_>)^2\Rbar_>^3\left( 3f_+(R_>)^2-2\right)} +\left(3\frac{f_T(\Rbar_>)^2}{f_+(\Rbar_>)^2}-1 \right)\left(9\frac{f_T(\Rbar_>)^2}{f_+(\Rbar_>)^2}+\frac13\right) \nonumber \\
    &-\frac{N^4\left(9f_+(\Rbar_>)^2-2 \right)+6\Rbar_>^2N^2f_+(\Rbar_>)^2\left(3f_+(\Rbar_>)^2-2\right)^2}{4f_+(\Rbar_>)^4\Rbar_>^6\left( 3f_+(\Rbar_>)^2-2\right)^3}~.
\end{align}

We insert these relations into the exact equations of motion at $\Rbar_<$ and $\Rbar_>$. At $\Rbar_<$ we find
\beq
\frac{f''_T(\Rbar_<)}{f_T(\Rbar_<)}+\frac{1}{\Rbar_<}\frac{f'_T(\Rbar_<)}{f_T(\Rbar_<)}+3f_+(\Rbar_<)^4-4f_+(\Rbar_<)^2+4f_T(\Rbar_<)^2-3f_T(\Rbar_<)^4 =0~, \label{eq:determining-equation-R<}
\eeq
and at $\Rbar_>$ we have
\beq
\frac{f''_T(\Rbar_>)}{f_T(\Rbar_>)}+\frac{1}{\Rbar_>}\frac{f'_T(\Rbar_>)}{f_T(\Rbar_>)}+3f_+(\Rbar_>)^4-4f_+(\Rbar_>)^2+4f_T(\Rbar_>)^2-3f_T(\Rbar_>)^4 =0~. \label{eq:determining-equation-R>}
\eeq
This system of equations can be solved (often numerically) for the two radii given specific values of $\kappa$ and $N$.

While the complete system is somewhat intractable, these equations simplify in certain cases. For $\Delta\Rbar>2$ the quantity $e^{-2\Delta\Rbar}<0.02$ may typically be ignored. In this case $f_T(\Rbar_{<,>})\to f_+(\Rbar_{<,>})/\sqrt{3}$ and the two equations completely decouple.

While each independent equation remains highly nontrivial, we make progress by applying our understanding of the potential. In Eq.~\eqref{eq:NoverKappa} when $\rbar=N/\kappa$ the maximum of the potential is at $V_N=0$. At any finite $R_>$ the particle cannot return to the origin without losing energy to friction. To provide this energy, we expect 
\beq
R_> >\frac{N}{\kappa}~.
\eeq

Using this value as a convenient dividing line, we parameterize the difference between $R_<, R_>$, and $N/\kappa$ as
\beq
\Rbar_<=\frac{N}{\kappa}-c_<~, \ \ \ \ \Rbar_>=\frac{N}{\kappa}+c_>~. \label{eq:R-bar-less-and-greater-simple-approximations-in-kappa}
\eeq
We expand the equations that determine the radii in the parameter $\kappa/N<1$. At leading order we find that the deviations are equal
\beq
c_<=\frac{1+\kappa^2}{4\kappa^2}~, \ \ \ \ c_>=\frac{1+\kappa^2}{4\kappa^2}~.
\eeq
The equation for $R_<$ cannot be trusted for small $\kappa$ because it predicts negative values when
\beq
\kappa<2N\left(1-\sqrt{1-\frac{1}{4N^2}} \right)~.
\eeq
The true values are, of course, always positive. This issue does not arise for $R_>$ or $\Delta\Rbar$.

These radius predictions imply that the difference in the radii
\beq
\Delta\Rbar=\frac{1+\kappa^2}{2\kappa^2}~, \label{eq:delta-R-bar-analytical-approximation}
\eeq
is independent of $N$, in very good agreement with the numerical results shown in Sec~\ref{sec:numerical-analysis}. Within this approximation we also see that $\Delta\Rbar>2$ for $\kappa<1/\sqrt{3}\approx0.58$, so we expect the results to be most accurate for $\kappa$ smaller than this value.

The average radius is 
\beq
\Rbar_a\equiv\frac12\left(\Rbar_>+\Rbar_< \right)=\frac{N}{\kappa}~. \label{eq:R-bar-average-analytical-approximation}
\eeq
This grows with $N$, but does not follow the $\kappa^{-2}$ growth of the Q-disk radius shown in Eq.~\eqref{eq:R-bar-approximation-formula}. The average radius also provides a way to estimate the height, or maximum amplitude, of the Q-ring. Equation~\eqref{eq:NoverKappa} implies that at leading order in $N/\kappa$ that
\beq
f_+(\kappa,\Rbar_a)=1~.
\eeq
Inserting this into the transition function form~\eqref{eq:Q-ring-transition-function} we find
\beq
f_T(\Rbar_a)=\frac{1}{1+2e^{-\Delta\Rbar}}~,\label{eq:RingMax}
\eeq
which can be used, with the estimate of $\Delta\Rbar$ in~\eqref{eq:R-bar-average-analytical-approximation}, to approximate the ($N$ independent) maximum height of the Q-ring profile.

We can also use the transitions function to approximate $Q$ and $E$. The leading order results are
\begin{align}
Q\approx&\,\frac{2\pi\phi_0^2\widetilde{\omega}f_+^2(\Rbar_a)\Rbar_a}{m_\phi^2-\omega_0^2}\frac{\Delta\Rbar-\ln2}{1-4e^{-2\Delta\Rbar}}+\mathcal{O}\left( e^{-\Rbar_a}\right)~, \label{eq:rotating-Q-approximation}\\
E\approx&\,\widetilde{\omega}Q+\pi\phi_0^2f_+^2(\Rbar_a)\frac{N^2}{\Rbar_a}\frac{\Delta\Rbar-\ln2}{1-4e^{-2\Delta\Rbar}}\nonumber\\
&+\pi\phi_0^2f_+^2(\Rbar_a)\frac{\Rbar_a}{2}\frac{1-16e^{-2\Delta\Rbar}\left(\Delta\Rbar-\ln2 \right)-16e^{-4\Delta\Rbar}}{\left( 1-4e^{-2\Delta\Rbar}\right)^3}+\mathcal{O}\left( e^{-\Rbar_a}\right)~.\label{eq:rotating-E-approximation}
\end{align}
These results provide close approximations of the exact energies and particle numbers for small $\kappa$ and large $N$, so that $\kappa/N$ is small, as shown in following section.

\section{Numerical Analysis} \label{sec:numerical-analysis}
The purpose of the analytical approximations introduced in Sec.~\ref{sec:Analytic} is to eliminate the need to numerically solve the nonlinear differential equations describing the profile, $f$ (at least for certain classes of solutions). Thus, these approximations are only useful provided they agree with the true values. In this section we compare our numerical results with the analytical approximations developed for Q-disks and Q-rings. We also use the numerical data derived in this section to identify qualitative trends and verify the differential relationship~\eqref{eq:GenDifRel} introduced in Sec.~\ref{sec:rotating-Q-balls}.

To determine the profile, $f$, of these configurations, we apply Mathematica's~\cite{Mathematica} finite element method to solve the equations of motion. This method avoids some difficulties encountered in other approaches, such as the double shooting method. To enforce the boundary condition at infinite radius, we employ a compactified radial coordinate given by
\beq
    y = \frac{\rbar}{1+\rbar/c}~,
\eeq
where $y=0$ when $\rbar = 0$ and $y$ increases monotonically with increasing $\rbar$, such that $y \to c$ (chosen to be a finite, positive number) as $\rbar \to \infty$. We use the resulting numerical solution for $f$ to compute numerical values of $R_<,\, R_>,\, Q,$ and $E$. 

For each $N$ and $\kappa$ we use the corresponding transition function\textemdash\eqref{eq:Q-disk-transition-function} or \eqref{eq:Q-ring-transition-function}\textemdash as an initial seed. We estimate $R_<$ and $R_>$ by numerically solving equations \eqref{eq:determining-equation-R<} and \eqref{eq:determining-equation-R>}. Using this approach, one may efficiently parallelize computations at various $N$ and $\kappa$. For some $\kappa$ and $N$, seed functions are not close enough for finite element method to robustly converge. In these cases we take neighboring (same $N$, similar $\kappa$) numerical solutions as the seed. By taking small steps in $\kappa$ between neighboring solutions, one may ``crawl'' through the more difficult parameter space. When using this method, however, simple parallelization is no longer possible. For larger $N$ this also requires taking increasingly small steps in $\kappa$.

\subsection{Profile Comparison}\label{sec:numerical-analyis_profile-comparison}
\begin{figure}[th]
  \centering
  \begin{subfigure}{0.49\textwidth}
    \centering
    \includegraphics[width=\linewidth]{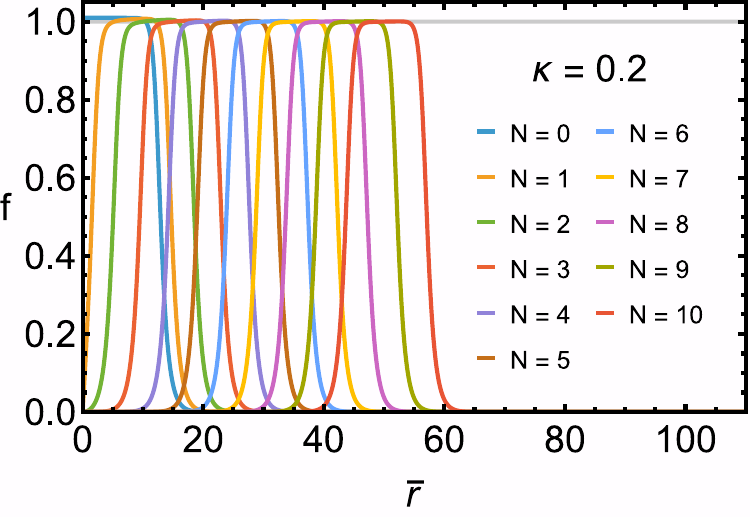}
  \end{subfigure}
  \hfill
  \begin{subfigure}{0.49\textwidth}
    \centering
    \includegraphics[width=\linewidth]{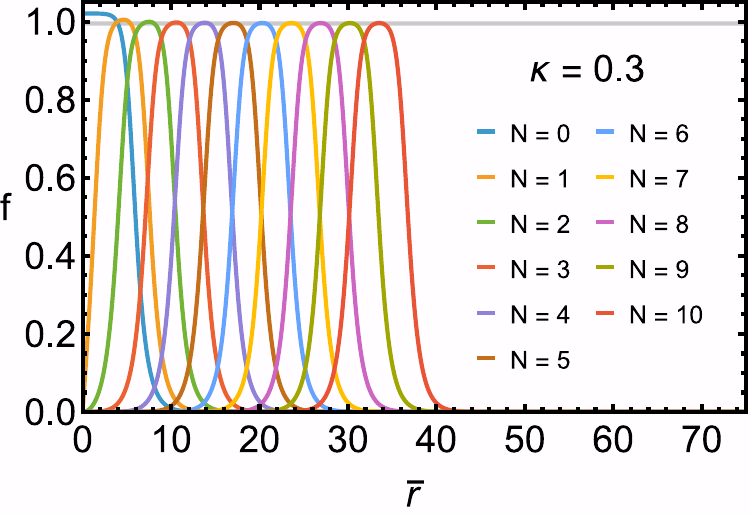}
  \end{subfigure}
  \begin{subfigure}{0.49\textwidth}
    \centering
    \includegraphics[width=\linewidth]{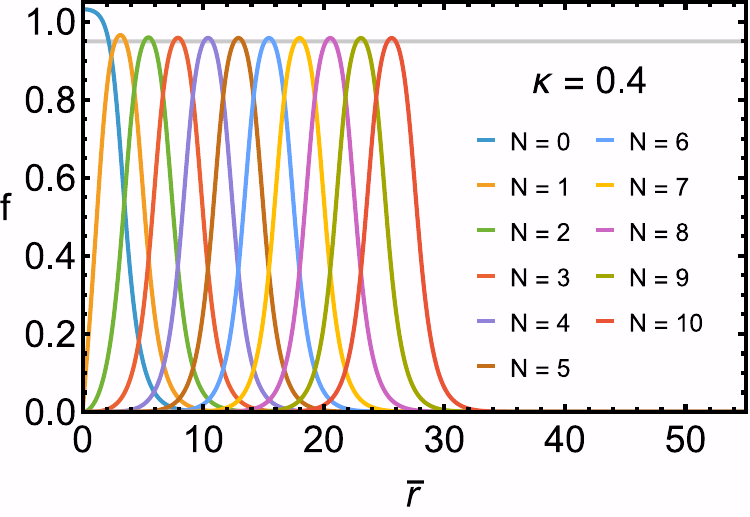}
  \end{subfigure}
  \hfill
  \begin{subfigure}{0.5\textwidth}
    \centering
    \includegraphics[width=\linewidth]{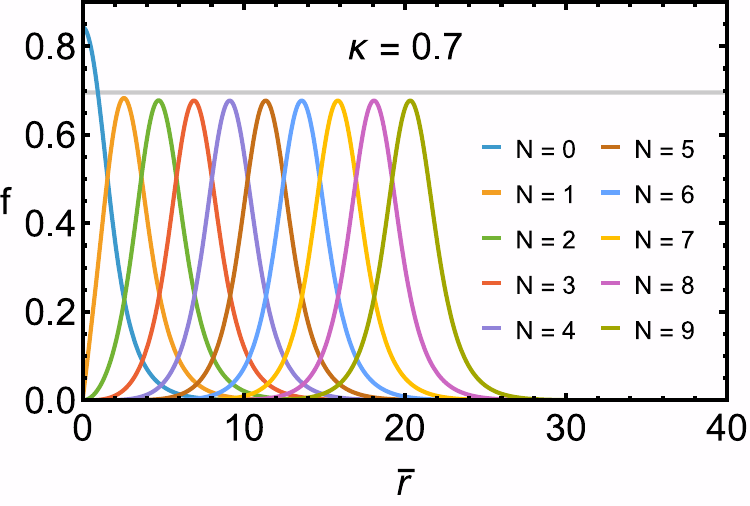}
  \end{subfigure}
  \caption{Numerical Q-ring profiles $f$ computed for various $N$ and $\kappa$. Horizontal and vertical scales of each plot differ. The horizontal gray line marks the analytical approximation for the Q-ring profile height.}
  \label{fig:profile-for-various-N-and-kappa}
\end{figure}

We begin by considering the Q-ring profiles. In Fig.~\ref{fig:profile-for-various-N-and-kappa} we find numerically computed profiles for several values of $N$ and $\kappa$. These profiles exhibit the qualitative features of the transition function form derived in Sec.~\ref{sec:Analytic}. 

For instance, the width ($\Delta \Rbar$) and maximum height of the Q-ring are independent of $N$ for each fixed $\kappa$, as suggested by \eqref{eq:delta-R-bar-analytical-approximation} and~\eqref{eq:RingMax}, respectively.  Moreover, $\Delta \Rbar$ increases as $\kappa$ decreases. For fixed $\kappa$, the radial location of the Q-ring's maximum amplitude point (as described by $\overline{R}_a$) increases linearly with $N$, in agreement with \eqref{eq:R-bar-average-analytical-approximation}. Furthermore, the amplitude at this maximum point decreases as $\kappa$ increases and is largely independent of $N$, as suggested by~\eqref{eq:RingMax}.

\begin{figure}[ht]
  \centering
  \includegraphics[width=0.6\linewidth]{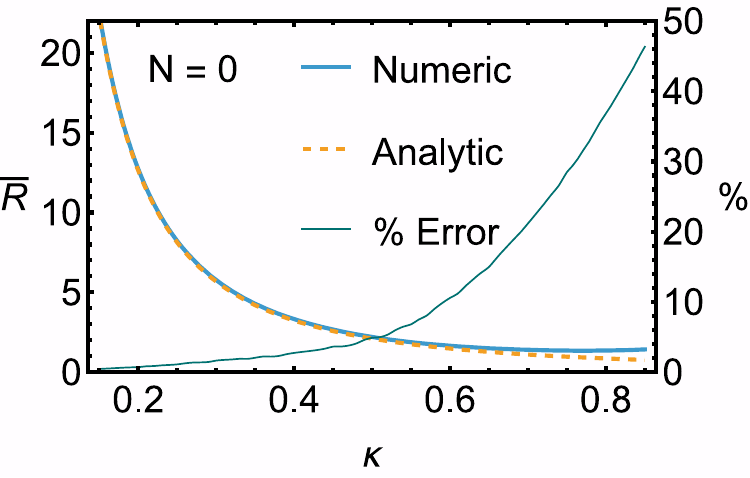}
  \caption{Comparison of Q-disk radius from numerical solutions (solid thick) and the analytical approximation (dashed). The solid thin curve shows the analytical approximation percent error (right axis).}
  \label{fig:R-bar-numerical-vs-analytical-approximation}
\end{figure}

The agreement between the transition function model and the numerical results is more than qualitative. Figure~\ref{fig:R-bar-numerical-vs-analytical-approximation} overlays the analytical model (dashed) prediction of the Q-disk radius~\eqref{eq:R-bar-approximation-formula} and the value extracted from the numerical results (solid thick) along with the percent error of the analytical results. For smaller values of $\kappa$ the two values are nearly identical. They begin to diverge as $\kappa$ increases, but the error remains below ten percent well beyond the thin-wall limit\textemdash up to around $\kappa = 0.6$.

\begin{figure}[ht]
  \centering
  \begin{subfigure}{0.49\textwidth}
    \centering
    \includegraphics[width=\linewidth]{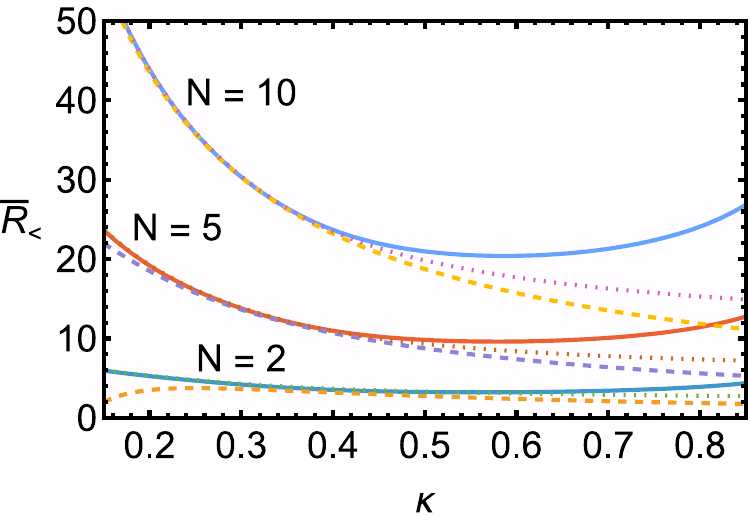}
  \end{subfigure}
  \hfill
  \begin{subfigure}{0.49\textwidth}
    \centering
    \includegraphics[width=\linewidth]{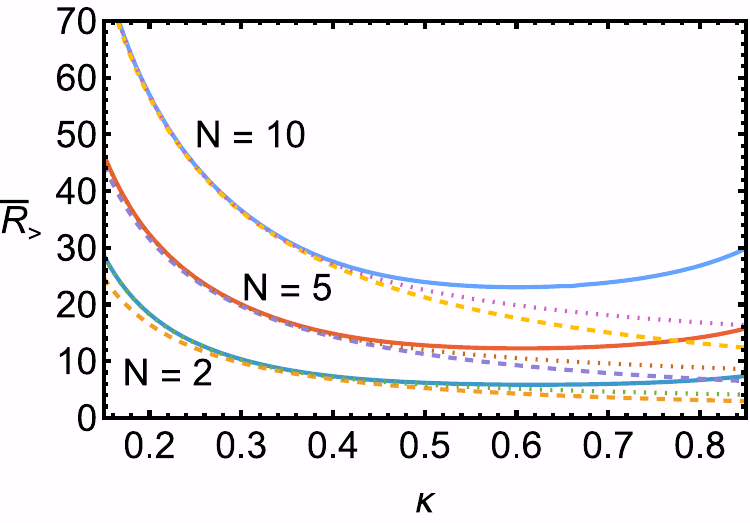}
  \end{subfigure}
  \caption{Numerical (solid line) values for the inner radius $\Rbar_{<}$ (left) and outer radius $\Rbar_{>}$ (right) compared with analytical approximations for $N=2,\, 5,\, 10$ Q-rings. Dotted curves indicate the full analytic approximation. Dashed lines are the leading order analytic approximation.} 
  \label{fig:R-bar-less-and-greater-approximation-comparison}
\end{figure}

Figure~\ref{fig:R-bar-less-and-greater-approximation-comparison} makes a similar comparison for rotating Q-rings with $N=2,\, 5,$ and 10. We consider both the inner ($\Rbar_<$ on left) and outer ($\Rbar_>$ on right) radii. The plot compares the numerical results (denoted with solid lines) to two analytical approximations. One (shown with dotted lines) assumes the transition function form and solves the two equations \eqref{eq:determining-equation-R<} and \eqref{eq:determining-equation-R>} to determine the radii. The second approximation drops terms of order $e^{-2\Delta\Rbar}$ that couple the equations and solves each to leading order in $\kappa/N$. The result \eqref{eq:R-bar-less-and-greater-simple-approximations-in-kappa} is plotted as the dashed lines in the figure. These plots show that for $\kappa\lesssim0.5$ the transition function model provides a very good approximation of the numerical results. Even the simple formula approximations are fairly accurate, becoming better as $N$ increases. 

\begin{figure}[ht]
  \centering
  \begin{subfigure}{0.52\textwidth}
    \centering
    \includegraphics[width=\linewidth]{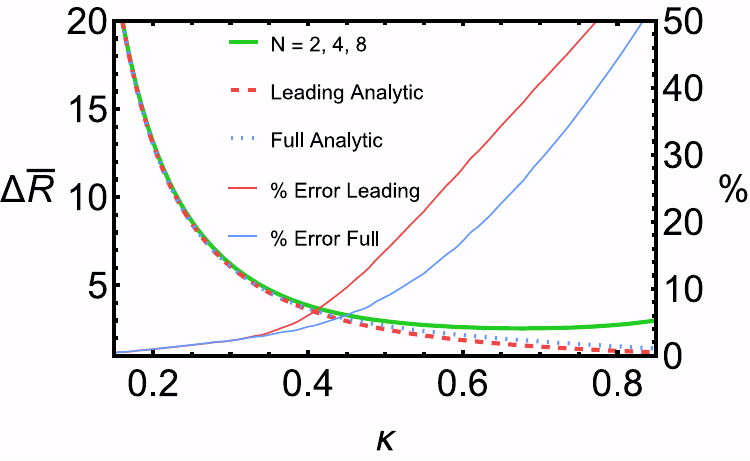}
  \end{subfigure}
  \hfill
  \begin{subfigure}{0.46\textwidth}
    \centering
    \includegraphics[width=\linewidth]{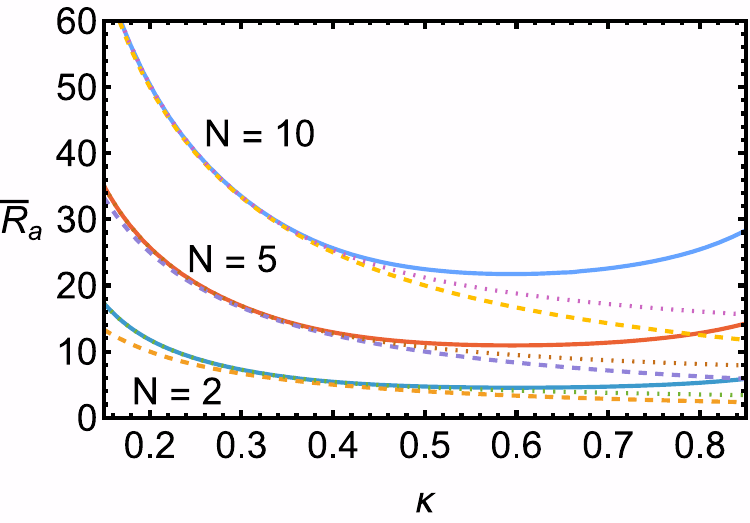}
  \end{subfigure}
  \caption{{\bf Left:} Numerical (solid) value of $\Delta \Rbar$  for $N=2, 4, 8$ as well as the full analytic approximation (dotted) and the leading order analytic approximation (dashed). {\bf Right:} Numerical value of $\Rbar_a$ (solid) for $N=2, 5, 10$ along with the full analytic approximation (dotted) the leading order analytical approximation (dashed).}
  \label{fig:Delta-R-bar-and-R-bar-a-approximation-comparison}
\end{figure}

If instead of focusing on the inner and outer radii we consider their average $\Rbar_a$ and difference $\Delta\Rbar$ we discover a somewhat surprising result. Figure~\ref{fig:Delta-R-bar-and-R-bar-a-approximation-comparison} plots the numerical values (solid line) of $\Delta\Rbar$ (left) and $\Rbar_a$ (right) as well as the two analytic approximations (dashed for full and dotted for leading) employed in the previous plots. While the agreement of these approximations of $\Rbar_a$ to the numerical values is similar to those of $\Rbar_{<,>}$, we find that the approximations for $\Delta \Rbar$ are considerably better. Perhaps the most interesting result is that the numerical $\Delta \Rbar$  appears to be $N$-independent, just as the leading order approximation \eqref{eq:delta-R-bar-analytical-approximation} suggests. We conjecture that this independence is not only approximately true to first order, but holds generally.

\subsection{Charge and Energy}
In this section we evaluate the transition function-based estimates for $Q$ and $E$ by comparing them to the numerical results. We also use the numerical data to display the general behavior of these quantities. To do so, we must choose numerical values for $\phi_0,\, m_\phi,$ and $\omega_0$, though the specific choice does not impact our conclusions. Without loss of qualitative generality, we chose $\phi_0 = 1,\, m_\phi = 1,$ and $\omega_0 = 0.5$. These values are used throughout this section. 

\begin{figure}[ht]
  \centering
  \includegraphics[width=0.6\linewidth]{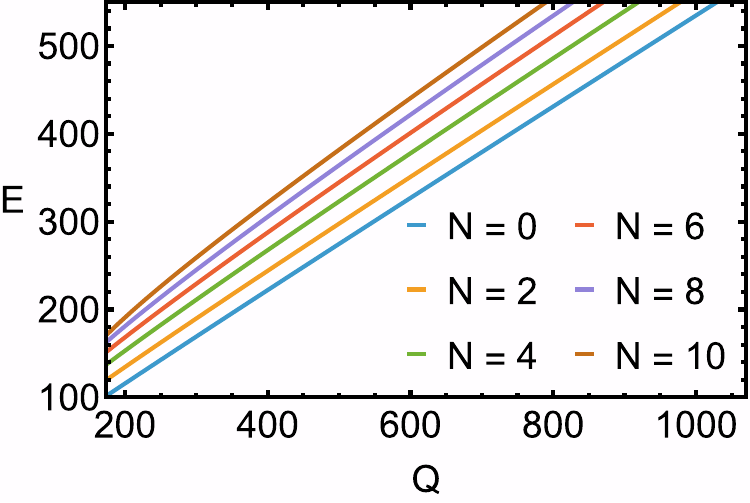}
  \caption{Q-disk ($N=0$) and Q-ring $(N>0)$ energies as a function of $Q$.}
  \label{fig:E-vs-Q}
\end{figure}

Figure~\ref{fig:E-vs-Q} displays the general relationship between $Q$ and $E$ for Q-disks ($N=0$) and Q-rings $(N>0)$. This figure makes obvious that (for fixed $N$) $E$ increases with increasing $Q$ and that (for fixed $Q$) $E$ increases as $N$ increases. Focusing on the lines of fixed $N$, the positive slope of each corresponds to the chemical potential\textemdash the energy increase of the system per unit charge added. In accordance with the differential relationship~\eqref{eq:general-differential-equation-with-dQ-and-dN}, $\widetilde{\omega}=\omega + N\Omega$ is the chemical potential and the combination of parameters that appears in the profile equation (within $\kappa$). This prediction is in exact agreement with the numerical solutions for each line of constant $N$.  

\begin{figure}[ht]
  \centering
  \begin{subfigure}{0.49\textwidth}
    \centering
    \includegraphics[width=\linewidth]{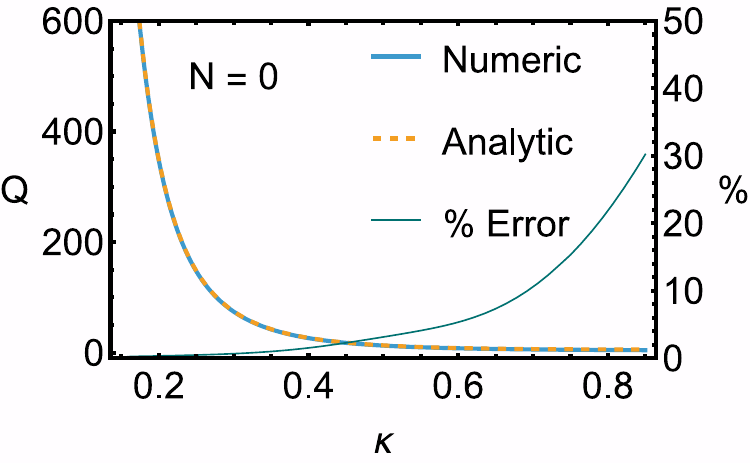}
  \end{subfigure}
  \hfill
  \begin{subfigure}{0.49\textwidth}
    \centering
    \includegraphics[width=\linewidth]{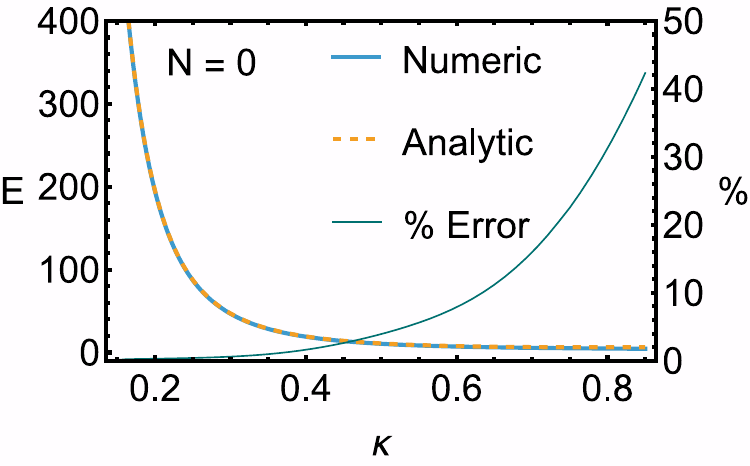}
  \end{subfigure}
  \caption{Comparison of the approximate analytic formulae (dashed) for $Q$ (left) and $E$ (right) with the numerical results (solid thick) for Q-disks. The percent error is given by the thin solid line (right axis).} 
  \label{fig:non-rotating-Q-and-E-approximation-comparison}
\end{figure}

We next consider the accuracy of the approximate formulae obtained in Sec.~\ref{sec:Analytic}. The Q-disk approximation for $Q$ is \eqref{eq:non-rotating-Q-approximation} and the approximation for $E$ is \eqref{eq:non-rotating-E-approximation}. Both of these approximations are given in terms of the Q-disk radius, $\Rbar$, which is approximated by \eqref{eq:R-bar-approximation-formula}. This was shown to be a good approximation of the numerical radius (defined by $f''(\Rbar)=0$) in Sec.~\ref{sec:numerical-analyis_profile-comparison}. Figure~\ref{fig:non-rotating-Q-and-E-approximation-comparison} summarizes the results, showing excellent agreement between the numerical and the approximate values of $Q$ and $E$. The percent error, shown with a thin solid line, for $Q$ ($E$) remains below 10\% up to $\kappa\approx0.7$ ($\kappa\approx0.65$). We see that the transition-based approximation of the Q-disk profile\textemdash which leads to the formulae for $Q$, $E$, and $\Rbar$\textemdash provides an accurate description of the solitons over a wide range of $\kappa$.

\begin{figure}[ht]
  \centering
  \begin{subfigure}{0.49\textwidth}
    \centering
    \includegraphics[width=\linewidth]{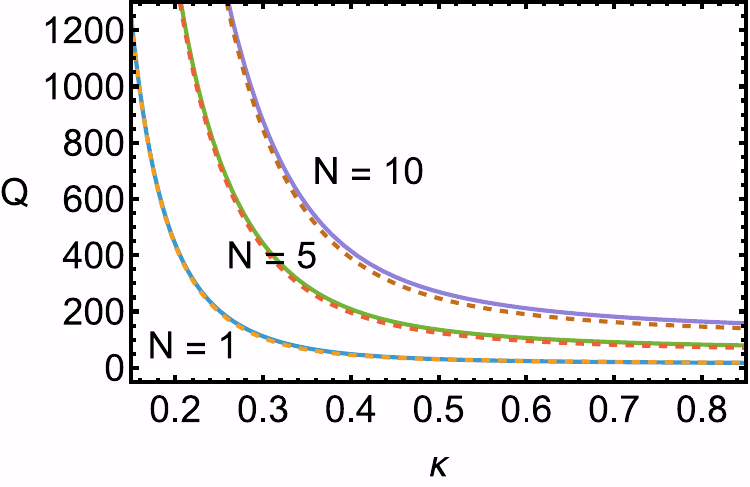}
  \end{subfigure}
  \hfill
  \begin{subfigure}{0.49\textwidth}
    \centering
    \includegraphics[width=\linewidth]{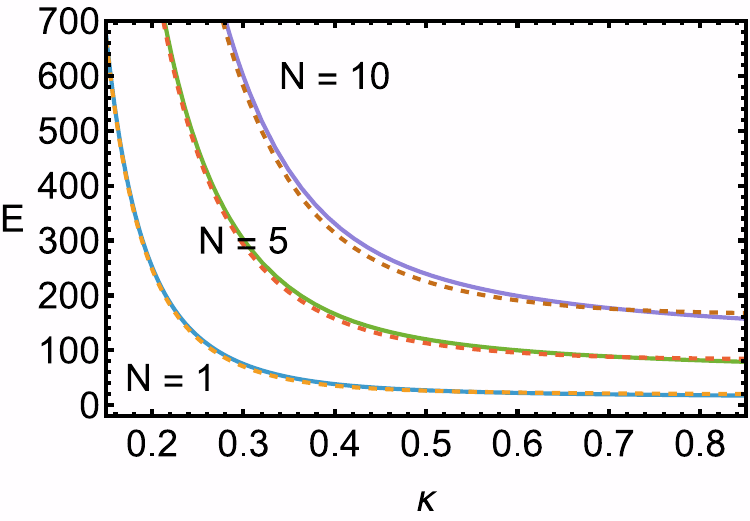}
  \end{subfigure}
  \caption{Comparison of the approximate analytic formulae (dashed) for $Q$ (left) and $E$ (right) with the numerical results (solid) for Q-rings with $N=1$, 5, and 10.} 
  \label{fig:rotating-Q-and-E-approximation-comparison}
\end{figure}

Finally, we consider the Q-ring approximations for $Q$ \eqref{eq:rotating-Q-approximation} and $E$ \eqref{eq:rotating-E-approximation}. These are given in terms of the average radis $\Rbar_a$ and the difference between the outer and inner radii $\Delta \Rbar$. The two radii are obtained by numerically solving the system of equations \eqref{eq:determining-equation-R<} and \eqref{eq:determining-equation-R>}, as in Sec.~\ref{sec:numerical-analyis_profile-comparison}. Figure~\ref{fig:rotating-Q-and-E-approximation-comparison} shows the excellent agreement between this approximation and the true numerical values for $Q$ and $E$ for a variety of $N$. We do not plot the percent error, but it is even better for Q-rings than what we found for Q-disks. For $Q$, the percent error remains below 11\% over the whole plotted $\kappa$ range and for all $N$ shown. The percent error for $E$ remains below 20\% for $N=1$ and below 10\% for all other $N$. As with the Q-disks, these results indicate that the analytic approximation provides an accurate description of these rotating solitons over wide ranges of $\kappa$ and $N$. 

\subsection{Differential Relationship}
In this section we investigate the quantities $\omega$ and $\Omega$. They were initially introduced as Lagrange multipliers to fix values of charge $Q$ and angular momentum $J$. They also appear in the differential relationship
\beq
dE = \omega\,dQ + \Omega\,dJ~,\label{eq:DiffAgain}
\eeq
which indicates they are conjugate variables to $Q$ and $J$. Just as each soliton has a real, physical value of $Q$ and $J$ they should also have a specific chemical potential $\omega$ and characteristic angular velocity $\Omega$.

The formulae for determining $Q$ and $J$ from a given soliton profile are straightforward. Equivalent methods for determining $\omega$ and $\Omega$ are not known. How can we find the values of $\omega$ and $\Omega$ for a given soliton? The equations that determine the profiles are specified by $N$ and $\widetilde{\omega} = \omega + N\Omega$ (through $\kappa$). For Q-disks $\Omega=0$ and $\widetilde{\omega} = \omega$ so the chemical potential is known once $\kappa$ is specified. Q-rings, however, are not so simple. For a given $N$ and $\widetilde{\omega}$, many $\omega$ and $\Omega$ satisfy
\beq
\omega=\widetilde{\omega}-N\Omega~.
\eeq
 
One way to determine these parameters is to turn to the differential relationship~\eqref{eq:DiffAgain}, which leads to
\begin{align}
    \omega &= \left(\frac{\partial E}{\partial Q}\right)_J~,&
    \Omega &= \left(\frac{\partial E}{\partial J}\right)_Q~,
\end{align}
where subscripts indicate a variable held constant in the partial derivative. One point is insufficient to compute a derivative, an open neighborhood about the point is required. Therefore, we numerically compute the profiles $f$ and their associated values $Q, J,$ and $E$ for many solitons in the neighborhood of the point in question. We then create an interpolation of these values in order to calculate derivatives. From these derivatives we extract values of $\omega$ and $\Omega$ that can then be compared with the $N$ and $\widetilde{\omega}$ associated with the solution at that point.

Since $Q$, $J$ and $E$ have an approximately exponential dependence on $\kappa$, we use $\ln Q,\, \ln J,$ and $\ln E$ to improve the accuracy of our interpolation. Specifically, we construct a two-dimensional interpolation function $F$ satisfying 
\beq
\ln E = F(\ln Q,\, \ln J)~.
\eeq 
It is difficult to pick values of $\kappa$ and $N$ such that $Q$ and $J$ lie on a regular grid, so we use radial basis function (RBF) interpolation, which is well-suited to unstructured data. We employ the standard RBF options in the \verb|scipy.interpolate| package~\cite{2020SciPy-NMeth}. Given the 2D interpolation function $F$, values for $\omega$ and $\Omega$ are computed as
\begin{align}
    \omega &= \left(\frac{\partial E}{\partial Q}\right)_J  = \frac{E}{Q}\frac{\partial F}{\partial (\ln Q)}~,&
    \Omega &= \left(\frac{\partial E}{\partial J}\right)_Q = \frac{E}{J}\frac{\partial F}{\partial (\ln J)}~.
\end{align}

\begin{figure}[ht]
  \centering
  \begin{subfigure}{0.49\textwidth}
    \centering
    \includegraphics[width=\linewidth]{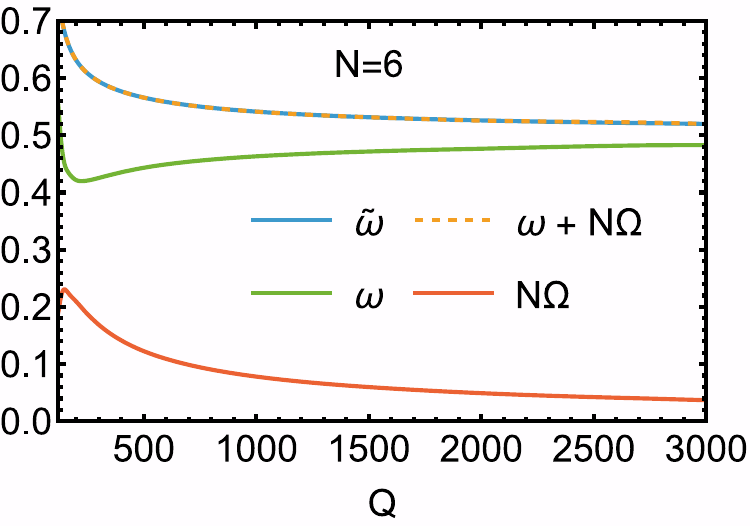}
  \end{subfigure}
  \hfill
  \begin{subfigure}{0.49\textwidth}
    \centering
    \includegraphics[width=\linewidth]{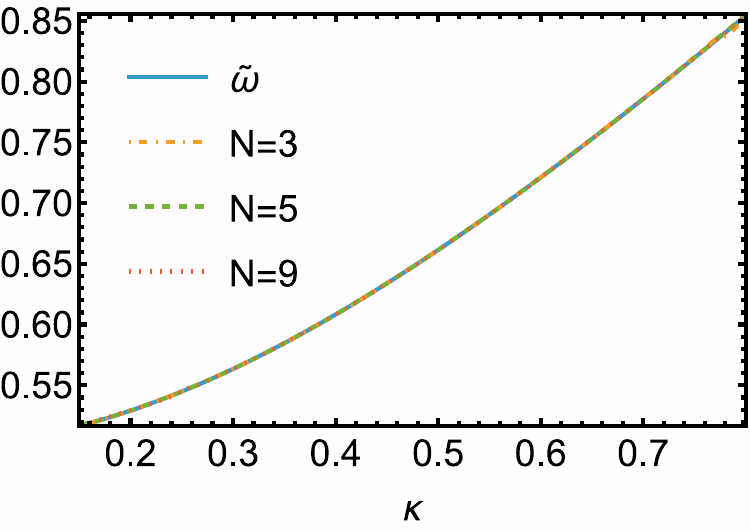}
  \end{subfigure}
  \caption{{\bf Left:} Comparison of values of $\omega$ and $\Omega$ extracted from the interpolation method (solid lines) with the true value of $\widetilde{\omega}$ (dashed line) as a function of $Q$. {\bf Right:} Comparison of the true value of $\widetilde{\omega}$ (solid line) with the interpolation results (dashed and dotted) for $N=3$, 5, and 9.} 
  \label{fig:omega-Omega-verification}
\end{figure}

In Fig.~\ref{fig:omega-Omega-verification} we compare these extracted values of $\omega$ and $\Omega$ with the constraint $\widetilde{\omega} = \omega + N\Omega$. In the left panel of the figure we consider the specific case of $N=6$. The solid lines denote the extracted values of $\omega$ (green) and $N\Omega$ (orange) as a function of $Q$. We sum these (solid blue) and  compare them with the true value of $\widetilde{\omega}$ (dashed yellow) for these numerical solutions. We see that the true values agree nearly exactly with those extracted from the interpolation procedure.

This panel of the figure is characteristic of what we find for all $N$ values. Thus, we can extract qualitative information about Q-rings from the plot. For instance, we see that for fixed $N$ as $Q$ increases (which also means that $J$ increases) that the characteristic angular velocity $\Omega$ decreases. This is a consequence of the increased average radius of the Q-ring. The distribution of large amplitude field is at larger radial values, so a slower angular velocity is needed to obtain the correct angular momentum. Such an interpretation suggests that other quantities, like momenta of inertia, might be defined and characterize these solitons, somewhat like the hydrodynamic aspects of Q-balls~\cite{Chen:2025tny}.

The right panel of Fig.~\ref{fig:omega-Omega-verification} simply plots the true value of $\widetilde{\omega}$ as a function of $\kappa$ as a solid blue line. This value is compared with the values of $\widetilde{\omega}$ generated by extracting $\omega$ and $\Omega$ from the interpolation function. This comparison is made for $N=3$, 5, and 9. In each case the agreement is remarkable, indicating the interpolation method is trustworthy over a range of $N$ values.
 
\section{Conclusion} \label{sec:conclusion}
In this article we derived the structure of rotating Q-balls in two and three spatial dimensions and elucidated the origin of their quantized angular momentum. Rather than assuming a particular rotating ansatz, we showed the solutions' form follows from minimizing the energy functional at fixed charge and angular momentum. We also showed the Lagrange multipliers $\omega$ and $\Omega$, which ensure constant $Q$ and $J$, have physical meaning as the chemical potential and the characteristic angular velocity, respectively. 

In addition, we developed approximate analytic descriptions of these solitons in two spatial dimensions, including expressions for their profiles, radii, charges, and energies. When compared to the numerical solutions, the approximations proved to be quite accurate over a broad range of parameter values. Depending on the application, these analytic results can be used instead of numerical solutions. We also used an interpolation of our numerical data to extract the values of $\omega$ and $\Omega$ for each soliton. A more detailed investigation of the rotational properties of these field configurations, such as a possible description of their moment of inertia, would be interesting. It is also natural to explore excited configurations that do not satisfy $J=NQ$.

Although our analysis focused on a simple class of solitons, the methods developed here may be applicable to more general scenarios. For instance, solitons that include multiple fields or an attractive force~\cite{Hamada:2024pbs} may lead to interesting modifications of our results. Our characterization of rotating solitons could also prepare the way for further investigation of their properties, including their stability~\cite{Gleiser:2005iq,Sakai:2007ft,Chen:2025oxo}, superradiance~\cite{Saffin:2022tub,Zhang:2025nqr}, oscillon modes~\cite{Martinez:2025ana}, and the effects of quantizing the scalar field~\cite{Xie:2023psz}. 

In addition, the two-dimensional solitons we described are easily extended to string-like solitons in three dimensions. Such objects have some similarities to the giant vortex strings~\cite{Dumitrescu:2025fme} of U(1) gauge theories, suggesting that the techniques developed in this work may prove useful in the study of gauge solitons.

Finally, while our explicit analytic treatment focused on the two-dimensional solitons, extending these methods to rotating Q-balls in three dimensions would be particularly interesting. This requires solving partial differential equations rather than ordinary differential equations, making numerical solutions substantially more challenging and increasing the potential value of analytic approximations. Many qualitative features discussed here are expected to persist, such as the increase in energy with angular momentum for fixed $Q$. It would be particularly interesting to confirm the differential relationships related to the characteristic angular velocity and the chemical potential for solitons in three spatial dimensions.

\acknowledgments
We are grateful to Eric Hirschmann for many helpful discussions. In addition, F.V. thanks Dan Homan for supervision at Denison University and the Departments of Physics and Astronomy at Brigham Young University and Denison University for their support and hospitality at different stages of this project. The work of C.B.V. and B.D. is supported in part by the National Science Foundation under grant No. PHY-2210067. F.V. received partial support from the College of Physical and Mathematical Sciences at Brigham Young University and from the Office of the Vice Chancellor for Research at the University of Wisconsin–Madison, with funding from the Wisconsin Alumni Research Foundation.

\appendix

\section{Full Analytical Approximation\label{app:FullAnApprox}}
In this appendix we extend the transition function approximation of the Q-disks and Q-rings. We find approximate solutions of the equations of motion in regions well before and after the transition. These three regions are joined together by requiring continuity in the profile and its first derivative.

\subsection{Q-disks} \label{appendix:non-rotating-analytical-model}

The Q-disk equation of motion~\eqref{eq:non-rotating-q-disk-equation-of-motion} is considered in three different regions: the exterior $(\rbar \gg \Rbar)$, interior $(\rbar \ll \Rbar)$, and transition $(\rbar \approx \Rbar)$ regions. The transition region derivation is found above in Sec.~\ref{sec:non-rotating-transition-function}.

In the exterior, since $f$ is small, we can expand the $dV/df$ term as
\begin{gather}
    f'' + \frac{1}{\rbar} f' + \left.\frac{dV}{df}\right|_{f = 0} + f \left.\frac{d^2 V}{df^2}\right|_{f = 0} \approx 0~,\\
    f'' + \frac{1}{\rbar} f' + (\kappa^2 - 1) f \approx 0~.
\end{gather}
The solutions to this equation are zeroth-order Bessel and Hankel functions. Choosing the option that satisfies $f(\rbar \rightarrow \infty) \approx 0$, the exterior ansatz becomes
\begin{align}
    f_>(\rbar) = c_> i H_0^{(1)}(i \sqrt{1 - \kappa^2} \rbar)~,
\end{align}
where $c_>$ is a real arbitrary constant to be determined later by matching.

The interior region $f$ reaches its maximum value. For $\kappa$ not too large this maximum is $f\approx f_+$ (see Fig.~\ref{fig:potential}). By expanding $dV/df$ around $f = f_+$ we find
\begin{gather}
    f'' + \frac{1}{\rbar} f' + \left.\frac{dV}{df}\right|_{f = f_+} + (f-f_+) \left.\frac{d^2 V}{df^2}\right|_{f = f_+} \approx 0~,\\
    f'' + \frac{1}{\rbar} f' - \alpha^2 (f - f_+) \approx 0~,
\end{gather}
where $\alpha$ is a $\kappa$-dependent constant given by
\begin{align}
    \alpha^2 = \frac{4}{3} \left(1 + 3 \kappa^2 + 2 \sqrt{1 + 3 \kappa^2}\right).
\end{align}
The solutions to this equation are also zeroth-order Bessel and Hankel functions. Choosing the one that is finite at the origin, we arrive at
\begin{align}
    f_<(\rbar) = f_+ + c_< J_0 (i \alpha \rbar)~.
\end{align}
We combine the solutions in each region for the complete analytical model
\begin{align}
	f(\rbar) =
	f_+
	\begin{cases} 
		1 + c_< J_0(i \alpha \rbar) & \textrm{for } \rbar < \Rbar_< \\
		\left[1 + 2 e^{2(\rbar - \Rbar)}\right]^{-1/2} & \textrm{for } \Rbar_< < \rbar < \Rbar_> \\
		c_> i H^{(1)}_0(i \sqrt{1 - \kappa^2} \rbar) & \textrm{for } \rbar > \Rbar_> \label{eq:full-profile-nonrot}
	\end{cases}~,
\end{align}
where we redefine the scaling constants to extract the overall scaling $f_+$ in accordance with our discussion in Sec.~\ref{sec:non-rotating-transition-function}. All that remains is to find $c_<$ and $c_>$.

Requiring that $f$ and $f'$ be continuous at $\rbar = \Rbar_<$ leads us to the condition
\begin{align}
	c_< = \frac{\left[1 + 2e^{2(\Rbar_< - \Rbar)}\right]^{-1/2} - 1}{J_0(i \alpha \Rbar_<)}~.
\end{align}
Meanwhile, the interior matching point is given by the equation
\begin{align}
	\alpha \frac{J_1(i \alpha \Rbar_<)}{J_0(i \alpha \Rbar_<)} = - \frac{2ie^{2(\Rbar_< - \Rbar)}}{1 + 2e^{2(\Rbar_< - \Rbar)} - \left[1 + 2e^{2(\Rbar_< - \Rbar)}\right]^{3/2}}~.
\end{align}
Unfortunately, there is no analytical solution to this equation, but one can easily find a value for $\Rbar_<$ using standard numerical methods.

Lastly, one can also find an equation for $c_>$ by enforcing the same requirements on $f$ and $f'$ at the exterior matching point $\Rbar_>$
\begin{align}
	c_> = \frac{\left[1 + 2e^{2(\Rbar_> - \Rbar)}\right]^{-1/2}}{i H_0^{(1)}(i \sqrt{1 - \kappa^2} \Rbar_>)}~.
\end{align}
Again, there is no simple solution for $\Rbar_>$, but numerical methods can easily be employed:
\begin{align}
	\frac{H^{(1)}_1 (i \sqrt{1 - \kappa^2} \Rbar_>)}{H^{(1)}_0 (i \sqrt{1 - \kappa^2} \Rbar_>)} = - \frac{2i}{\sqrt{1 - \kappa^2}\left[2 + e^{-2(\Rbar_> - \Rbar)}\right]}~.
\end{align}

\subsection{Q-rings}
\label{appendix:rotating-analytical-model}

The full analytical approximation for Q-rings follows a similar sequence of steps as Sec~\ref{appendix:non-rotating-analytical-model}. This time, however, we seek to solve the $N$-dependent equation of motion in Eq.~\eqref{eq:rotating-q-disk-equation-of-motion}.

To find the exterior ansatz, we consider the limit where $\rbar \gg \Rbar_>$. Therefore, we can expand $dV/df$ around $f = 0$, obtaining
\begin{gather}
    f'' + \frac{1}{\rbar} f' - \frac{N^2}{\rbar^2} f + \left.\frac{dV}{df}\right|_{f = 0} + f \left.\frac{d^2V}{df^2}\right|_{f = 0} = 0,\\
    f'' + \frac{1}{\rbar} f' - \frac{N^2}{\rbar^2} f + (\kappa^2 - 1)f = 0~. \label{eq:eom-exterior}
\end{gather}
The solutions to this equation are $N$$^{\textrm{th}}$ order Bessel and Hankel functions. By considering that $f(\rbar = 0) = 0$ and that $f$ must be real, one finds that the solution must take the following form
\begin{align}
    f_> (\rbar) = c_> i^{N+1} H_N^{(1)} \left(i \sqrt{1 - \kappa^2} \rbar\right)~.
\end{align}

For the interior region, we consider the limit of $\rbar \ll \Rbar_<$. Since $f \approx 0$ in this limit, we may once again expand $dV/df$ around $f = 0$, yielding the same equation as in the previous step. In this case, however, one finds that the only sensible solution is 
\begin{align}
    f_<(\rbar) = c_< (-i)^N J_N \left(i \sqrt{1 - \kappa^2} \rbar\right)~.
\end{align}
Therefore, our complete ansatz is
\begin{align}
    f(\rbar) =
    \begin{cases}
        c_< (-i)^N J_N \left(i \sqrt{1 - \kappa^2} \rbar\right), & 0 < \rbar < \Rbar_<,\\
        \left[\left(1 + 2 e^{-2(\rbar - \Rbar_<)}\right)\left(1 + 2 e^{2(\rbar - \Rbar_>)}\right)\right]^{-1/2}, & \Rbar_< < \rbar < \Rbar_>,\\
        c_> i^{N+1} H_N^{(1)} \left(i \sqrt{1 - \kappa^2} \rbar\right), & \Rbar_> < \rbar < \infty~.
    \end{cases}
\end{align}

We find the matching points and scaling constants by requiring that $f$ and $f'$ are continuous at $\rbar = \Rbar_<$ and $\rbar = \Rbar_>$. First, if we require that $f$ be continuous at $\rbar = \Rbar_<$, we get that $c_<$ must be
\begin{align}
    c_< = \frac{\left[\left(1 + 2 e^{-2(\Rbar_< - \Rbar_<)}\right)\left(1 + 2 e^{2(\Rbar_< - \Rbar_>)}\right)\right]^{-1/2}}{(-i)^N J_N \left(i \sqrt{1 - \kappa^2} \Rbar_<\right)}~.
\end{align}
Meanwhile, if we use this result and require that $f'$ also be continuous at $\rbar = \Rbar_<$, we find that
\begin{gather}
    \frac{i \sqrt{1 - \kappa^2}}{J_N \left(i \sqrt{1 - \kappa^2} \Rbar_<\right)} \left[J_{N-1} \left(i \sqrt{1 - \kappa^2} \Rbar_< \right) - J_{N+1} \left(i \sqrt{1 - \kappa^2} \Rbar_< \right)\right] \nonumber \\
    = 4 \left[e^{-2(\Rbar_<-\Rbar_<)}-e^{2(\Rbar_<-\Rbar_>)}\right]\left[\left(1 + 2 e^{-2(\Rbar_< - \Rbar_<)}\right)\left(1 + 2 e^{2(\Rbar_< - \Rbar_>)}\right)\right]^{-1}~.
\end{gather}
One can simplify this by considering that the ratio of Bessel functions on the left is approximately equal to $-2i$ for $\Rbar_< \gtrsim 4$, yielding the following equation for $\Rbar_<$:
\begin{align}
    \sqrt{1 - \kappa^2} \left(1 + 2 e^{-2(\Rbar_< - \Rbar_<)}\right)\left(1 + 2 e^{2(\Rbar_< - \Rbar_>)}\right) = 2 \left(e^{-2(\Rbar_<-\Rbar_<)} - e^{2(\Rbar_<-\Rbar_>)}\right).
\end{align}
Then, if we require that $f$ be continuous at $\rbar = \Rbar_>$, we obtain the following expression for $c_>$:
\begin{align}
    c_> = \frac{\left[\left(1 + 2 e^{-2(\Rbar_> - \Rbar_<)}\right)\left(1 + 2 e^{2(\Rbar_> - \Rbar_>)}\right)\right]^{-1/2}}{i^{N+1} H_N^{(1)} \left(i \sqrt{1 - \kappa^2} \Rbar_>\right)}.
\end{align}
Similarly, requiring that $f'$ be continuous at $\rbar = \Rbar_>$ yields the following equation:
\begin{gather}
    \frac{i \sqrt{1 - \kappa^2}}{H^{(1)}_{N} \left(i \sqrt{1 - \kappa^2} \Rbar_>\right)} \left[H^{(1)}_{N-1} \left(i \sqrt{1 - \kappa^2} \Rbar_> \right) - H^{(1)}_{N+1} \left(i \sqrt{1 - \kappa^2} \Rbar_> \right)\right] \nonumber \\
    = 4 \left[e^{-2(\Rbar_>-\Rbar_<)}-e^{2(\Rbar_>-\Rbar_>)}\right]\left[\left(1 + 2 e^{-2(\Rbar_> - \Rbar_<)}\right)\left(1 + 2 e^{2(\Rbar_> - \Rbar_>)}\right)\right]^{-1}.
\end{gather}
In this case, one finds that the ratios of Hankel functions on the left (excluding the $i \sqrt{1 - \kappa^2}$) is approximately equal to $+2i$ for $\Rbar_> \gtrsim 8$, arriving at the following equation for $\Rbar_>$:
\begin{align}
    - \sqrt{1 - \kappa^2} \left(1 + 2 e^{-2(\Rbar_> - \Rbar_<)}\right)\left(1 + 2 e^{2(\Rbar_> - \Rbar_>)}\right) = 2 \left(e^{-2(\Rbar_>-\Rbar_<)}-e^{2(\Rbar_>-\Rbar_>)}\right).
\end{align}

\bibliographystyle{JHEP}
\bibliography{BIB.bib}

\end{document}